\documentclass[fleqn]{2026SCGE}
\usepackage{bm}
\usepackage{algorithm}
\usepackage{hyperref}

\usepackage{threeparttable}
\usepackage{booktabs}
\usepackage{color}
\usepackage{graphicx}
\usepackage{subfigure}
\usepackage{longtable}
\usepackage{supertabular}
\usepackage{bm}
\usepackage{amsmath}
\usepackage{amssymb}

\begin{document}
\ensubject{subject}

\ArticleType{Article}
\SpecialTopic{SPECIAL TOPIC: }
\Year{2026}
\Month{January}
\Vol{69}
\No{1}
\DOI{??}
\ArtNo{000000}
\ReceiveDate{January 11, 2023}
\AcceptDate{April 6, 2023}

\title{Nucleosynthesis of Pop III and Fe-enriched Pop II Pair-Instability Supernovae}

\author[1,2]{Wenyu Xin}{{xinwenyu16@mails.ucas.ac.cn}}%
\author[3]{Ken'ichi Nomoto}{}
\author[4,5]{Chun-Ming Yip}{}  
\author[1,2]{Xianfei Zhang}{{zxf@bnu.edu.cn}}%
\author[6]{\\Qian-Fan Xing}{}
\author[1,2]{Shaolan Bi}{}
\author[6,7]{Gang Zhao}{}

\AuthorMark{Xin W. Y.}
\AuthorCitation{Xin W. Y., Nomoto K., Yip C. M., et al}

\address[1]{Institute for Frontiers in Astronomy and Astrophysics,
Beijing Normal University, Beijing 102206, People's Republic of China}
\address[2]{School of Physics and Astronomy, Beijing Normal University, Beijing 100875, People's Republic of China}
\address[3]{Kavli Institute for the Physics and Mathematics of the Universe (WPI), The University of Tokyo Institutes for Advanced Study, \\
The University of Tokyo, Kashiwa, Chiba 277-8583, Japan}
\address[4]{Department of Physics and Institute of Theoretical Physics, The Chinese University of Hong Kong,\\
Shatin, N.T., Hong Kong S.A.R., People's Republic of China}
\address[5]{GSI Helmholtzzentrum f\"ur Schwerionenforschung, Planckstra{\ss}e 1, D-64291 Darmstadt, Germany}
\address[6]{Key Laboratory of Optical Astronomy, National Astronomical Observatories, \\
Chinese Academy of Sciences (CAS), Beijing 100101, People's Republic of China}
\address[7]{School of Astronomy and Space Science, University of Chinese Academy of
Sciences, Beijing 100101, People's Republic of China}


\abstract{Recently discovered very metal-poor (VMP) star LAMOST J1010+2358 shows
a peculiar abundance pattern that is remarkably well fit by a Pop III pair-instability
supernova (PISN) of $\simeq 260$ M$_\odot$. Motivated by this,
we investigate the nucleosynthetic characteristics of Pop III and Pop II PISNe
to provide theoretical constraints for future observations.
This paper is divided into two parts.
First, we explore the evolution and nucleosynthesis of Pop III PISNe
with initial masses of 130 - 300 M$_\odot$.
Our main aim is to investigate how the uncertainty in
$^{12}$C$(\alpha,\gamma)^{16}$O and $^{16}$O+$^{16}$O reaction rates
affect their explosion properties and nucleosynthesis.
We find that the yields of odd-$Z$ elements are particularly sensitive
to the $^{12}$C$(\alpha,\gamma)^{16}$O rate,
while the production of Fe-peak elements shows significant sensitivity
to both the rates.
Second, we investigate the nucleosynthetic features of Pop II PISNe
formed in gas enriched exclusively by Pop III PISN ejecta.
By employing a time-dependent convection model during the explosion,
we demonstrate that metal enrichment increases opacity and
triggers vigorous convective mixing.
This hydrodynamic effect significantly enhances the explosion energy
and $^{56}\text{Ni}$ production.
Consequently, Pop II PISNe exhibit distinct chemical signatures,
including a weaker odd-even effect and enhanced Zn-Ge production,
providing unique diagnostics for identifying PISN remnants
in the early Universe.}


\maketitle


\begin{multicols}{2}

\section{Introduction} \label{sec:intro}

The mechanism of pair-instability supernova (PISN) was first proposed
in the 1960s \cite{1964ApJS....9..201F, 1967PhRvL..18..379B}.
Massive stars with the Zero-Age Main-Sequence (ZAMS) masses of
$M ({\rm ZAMS}) \simeq$ 140 - 300 M$_\odot$ are expected to develop He
cores of masses $M ({\rm He}) \simeq$ 65 - 140 M$_\odot$,
which eventually undergo thermonuclear explosions triggered
by an electron-positron pair-instability \cite{2001ApJ...550..890B}.
The detailed calculations of nucleosynthesis of Pop III PISNe
from refs. UN02 \cite{2002ApJ...565..385U},
HW02 \cite{2002ApJ...567..532H}, and TYU18 \cite{Takahashi_2018_1} predicted that 
the post-He burning phases of PISNe progenitors proceed under lower densities
than in less massive stars.
Consequently, the neutron excess in PISNe is insufficient to produce
significant amounts of odd-charge elements.
Because of the immense energy released by explosive burning,
PISNe completely disrupt the progenitors without leaving black holes and
eject large amounts of metals, especially $^{56}$Ni. 

PISNe are difficult to be observed directly in the local Universe
because their progenitors require to have low metallicities (Z$_\odot$/3)
to avoid significant mass loss via stellar winds, which would otherwise reduce
the He core mass \cite{2007AA...475L..19L, 2019ApJ...887...72L}.
Recent observations have provided indirect evidence.
Ref. \cite{2023Natur.618..712X} reported that the peculiar abundance pattern of 
the very metal-poor (VMP) star LAMOST J1010+2358 closely matches
the nucleosynthetic yields predicted for a Pop III PISN with $M ({\rm ZAMS}) =$ 260 M$_\odot$.
This discovery seems to provide some evidence for the existence of PISNe,
although some alternative interpretations (e.g., core-collapse supernova, hypernova,
or complex enrichment scenarios) are proposed by further observations and simulations
\cite{2024MNRAS.527.4790J, 2024OJAp....7E..66T, 2024OJAp....7E..83J}. 
Furthermore, observations from refs.
\cite{2022ApJ...925..111I, 2022ApJ...937...61Y, 2024ApJ...976..122N}
indicate high Fe abundances in extremely metal-poor galaxies
and high redshift objects, which provide clues to a significant contribution
from PISN ejecta in the early Universe. These observations motivate a
re-examination of Pop III PISNe’s role in early chemical enrichment.

While core-collapse supernovae, faint supernovae, and hypernovae are generally considered
the dominant sources of early metal enrichment \cite{2013ARAA..51..457N, Kobayashi_2020},
the detection of possible PISN signatures suggests that Pop III PISNe may have
played a non-negligible role. To explore the consequences of such enrichment,
we investigate the evolutionary and nucleosynthetic
differences between primordial Pop III PISNe and second-generation (Pop II) PISNe.
Furthermore, recent JWST observations of galaxies at $z > 10$ have 
revealed a rapid metal enrichment and substantial, highly localized chemical 
inhomogeneities \cite{2024ApJ...976..122N, 2026arXiv260608782N, 2026ApJ...997...86U}.
To provide a theoretical baseline for such highly localized chemical inhomogeneities
uncovered by JWST, we consider an idealized but physically motivated scenario: 
massive Pop II stars forming promptly in a highly localized region enriched solely
by Pop III PISNe, before being contaminated by other supernova products occurs.
The PISNe formed in such an Fe-enriched,
neutron-deficient extreme region may produce peculiar abundance patterns.
Such a model serves as a clean diagnostic tool to trace the localized chemical
evolution history for the future JWST observations.

Previous studies of Pop II PISNe by KYL14 \cite{2014AA...566A.146K} and UN24
\cite{2024ApJ...961..146U} have established key nucleosynthetic characteristics.
They found that metal-enriched PISNe retain a similar nucleosynthetic pattern
but a slightly weaker odd-even effect than in Pop III counterparts.
However, their nucleosynthesis calculations rely on post-processing and
neglect convective mixing during explosive burning,
despite that recent hydrodynamic simulations have shown that PISNe develop
extensive convective regions during the explosion phase
\cite{Farmer_2019, 2019ApJ...882...36M, 2020MNRAS.493.4333R, 
2022ApJ...937..112F, 2023RAA....23a5014X, 2026RAA....26g5011X}.
In contrast, our models include a treatment of explosive convective mixing based on the
time-dependent convection formalism developed by \cite{Farmer_2019, 2019ApJ...882...36M},
which accounts for the finite timescale of convective growth and decay during the dynamic explosion.

Addressing these issues, this paper is organized to answer whether Fe-enriched Pop II PISNe
(hereafter Fe-enriched PISNe) produce distinguishable signatures from Pop III PISNe.
In Section \ref{sec:model}, we describe our stellar models and some assumptions.
Then we divided this paper into two main parts:
In Part I (Section \ref{sec:popIII}), we present Pop III PISNe models and investigate
how the uncertainty in $^{12}$C$(\alpha,\gamma)^{16}$O and
$^{16}$O+$^{16}$O reaction rates affect the explosion properties and nucleosynthesis of PISNe.
In Part II (Section \ref{sec:popII}), we present our calculations of Fe-enriched
PISNe with explosive convective mixing and
discuss the implications for chemical enrichment.
In Section \ref{sec:summary}, we summarize this paper and show our conclusions.

\section{Stellar Models and Assumptions} \label{sec:model}

\subsection{Code and Assumptions}

All stellar evolution and nucleosynthesis calculations presented in this work
were performed using the Modules for Experiments in Stellar Astrophysics 
(MESA, version 15140; \cite{2011ApJS..192....3P, 2013ApJS..208....4P, 
2018ApJS..234...34P, 2019ApJS..243...10P})
\footnote{The inlist files used in this study are uploaded to
DOI:\href{https://zenodo.org/records/20609738}{10.5281/zenodo.20609738}. }.

For hydrostatic evolutionary phases, we employ a nuclear reaction network consisting
of 128 isotopes (\texttt{mesa\_128.net}),
while a larger 330-isotope network (\texttt{mesa\_330.net})
is adopted to capture the detailed nucleosynthesis during explosive burning.
The specific isotopes included in each network are listed in Table~\ref{tab:nuclear}.
Most thermonuclear reaction rates are taken from the latest JINA REACLIB compilation
\cite{2010ApJS..189..240C}, with the exception of the $^{12}$C$(\alpha, \gamma)^{16}$O
and $^{12}$C+$^{12}$C reaction rates. 
For $^{12}$C+$^{12}$C reaction rate, we combines the total rate from CF88
\cite{CAUGHLAN1988283} and the branching ratios of
$\alpha$, p, n channels from ref. \cite{1976NuPhA.265..153D}.
The choice of $^{12}$C$(\alpha, \gamma)^{16}$O will be discussed in Section \ref{sec:c12a}.
Weak interaction rates are from the tabulations from refs.
\cite{2000NuPhA.673..481L, ODA1994231, 1985ApJ...293....1F}.

Convective instability is determined using the Ledoux criterion.
For convective mixing during hydrostatic phases,
we adopt the Mixing-Length Theory (MLT) with a mixing-length parameter
$\alpha_{\rm mlt}$ = 2.0, which is defined as $\alpha_{\rm MLT}=\frac{\lambda}{H_p}$.
Semiconvective mixing is treated following \cite{1983AA...126..207L}
with an efficiency parameter $\alpha_{sc} =$1.0.
Convective boundary mixing is modeled via exponential overshooting with
parameters $f_0 = 0.005$ and $f = 0.01$,
which can soften convective boundaries and allow convective
regions to expand against steep composition gradients
\cite{2019ApJ...882...36M}.
The convection mixing during the explosion is discussed in Section
\ref{sec:TDC} in detail.
Since Pop III massive stars are metal-free,
wind mass loss is neglected throughout their evolution.

\begin{table}[H]
\centering
\caption{Isotopes included in this work before explosion (\texttt{mesa\_128.net})
and during the explosive burning (\texttt{mesa\_330.net})}
\label{tab:nuclear}
\begin{tabular}{cccccc}  
\toprule
Element & \multicolumn{2}{c}{$A$} & Element & \multicolumn{2}{c}{$A$} \\
\midrule
Isotopes & 128       & 330       & Isotopes & 128       & 330       \\
\midrule
n       & 1         & 1         & Cl      & 35-37     & 31-45      \\
H       & 1-2       & 1-3       & Ar      & 35-38     & 32-46     \\
He      & 3-4       & 3-4       & K       & 39-43     & 35-49     \\
Li      & 7         & 6-7       & Ca      & 39-44     & 36-49     \\
Be\tnote{1}      & 7-10      & 7-9       & Sc      & 43-46     & 40-51     \\
B       & 8         & 8-11      & Ti      & 44-48     & 41-53    \\
C       & 12-13     & 11-14     & V       & 47-51     & 43-55     \\
N       & 13-15     & 12-15     & Cr      & 48-57     & 44-58      \\
O       & 14-18     & 13-20     & Mn      & 51-56     & 46-59   \\
F       & 17-19     & 17-21     & Fe      & 52-58     & 47-64       \\
Ne      & 18-22     & 17-24     & Co      & 55-60     & 50-65     \\
Na      & 21-24     & 19-27     & Ni      & 59-62     & 51-66      \\
Mg      & 23-26     & 20-29     & Cu      & 60-64     & 55-67     \\
Al      & 25-28     & 22-31     & Zn      & $\cdots$  & 57-70   \\
Si      & 27-30     & 23-34     & Ga      & $\cdots$  & 61-73       \\
P       & 30-32     & 27-38     & Ge      & $\cdots$  & 62-76     \\
S       & 31-34     & 28-42     &         &           &           \\
\bottomrule
\end{tabular}
\begin{tablenotes}
\footnotesize
\item[1] $^{8}$Be is not included in \texttt{mesa\_128.net}.
\end{tablenotes}
\end{table}

\subsection{$^{12}$C$(\alpha, \gamma)^{16}$O reaction} \label{sec:c12a}

$^{12}$C$(\alpha, \gamma)^{16}$O reaction rate is active near the end of 
He burning and converts $X$($^{12}$C) to $X$($^{16}$O) in the core.
It is only $\sim$ 300 keV for the typical Gamow energy of this reaction
corresponding to the temperature during the He burning phase. Unfortunately,
the cross-section of $^{12}$C$(\alpha, \gamma)^{16}$O reaction
decreases significantly below this energy range,
reaching 10$^{-17}$ barns - far below the sensitivity
of even the most advanced current nuclear measurement techniques
\cite{2020PhRvL.124p2701S}.   
Because the cross sections at higher energies are complicated by
the interference from other excited states of $^{16}$O,
the extrapolation down to the astrophysically relevant energies is
highly desirable and challenging \cite{2017RvMP...89c5007D}.

\begin{figure}[H]
\centering
\begin{minipage}[c]{0.48\textwidth}
\includegraphics [width=80mm]{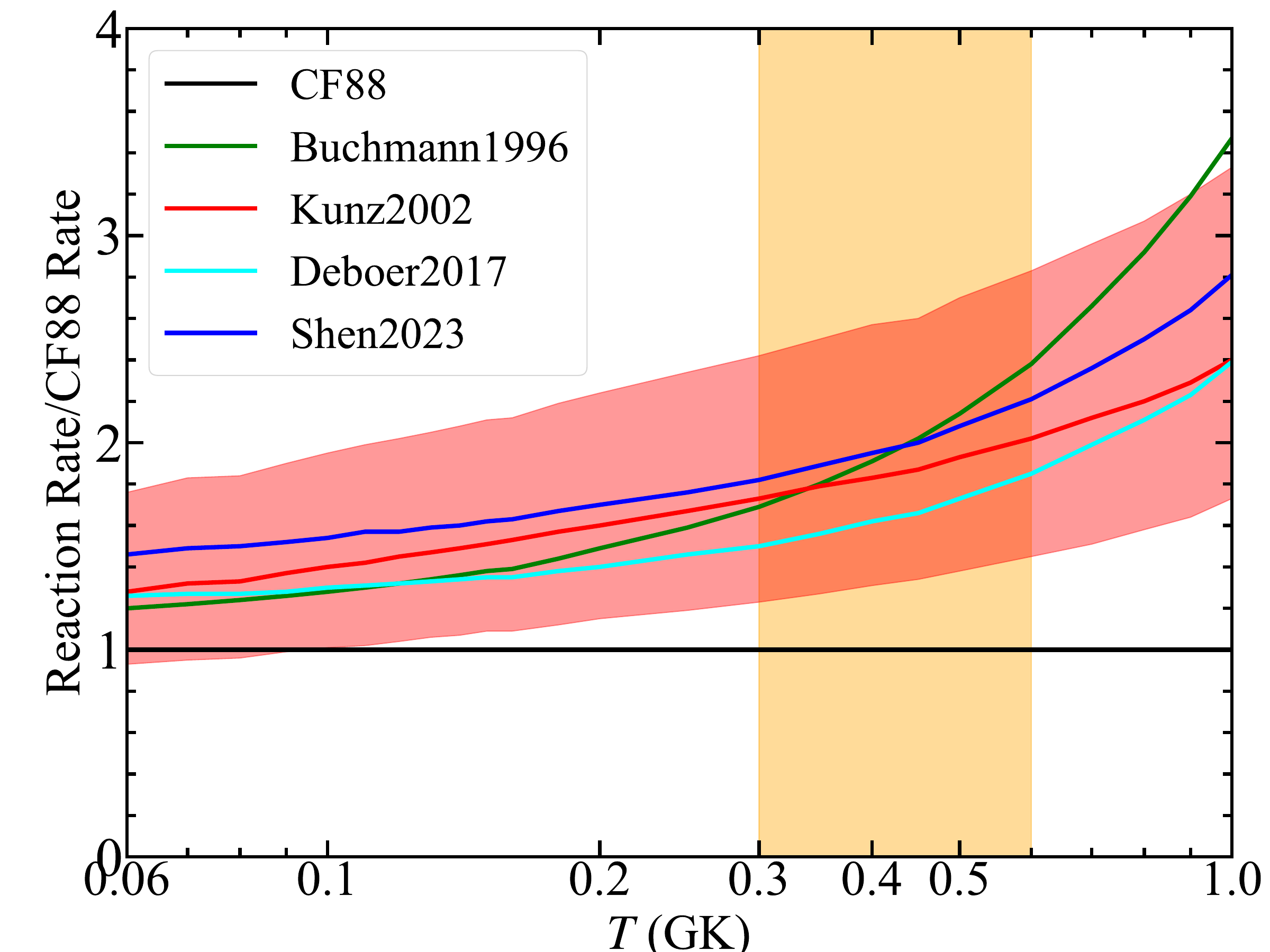}
\end{minipage}%
\caption{Comparison of the $^{12}$C$(\alpha, \gamma)^{16}$O reaction rates
in different references. The ratios between the rates from refs. 
\cite{1996ApJ...468L.127B, 2002ApJ...567..643K, 2017RvMP...89c5007D, 2020PhRvL.124p2701S}
and the CF88 rate are shown as a function of the temperature.
The uncertainty of the rate from ref. \cite{2002ApJ...567..643K} by
$\pm$ 1 $\sigma$ is shaded in red.
The temperature range relevant to the core helium burning is shaded in orange.
\label{fig:c12a_rate}}
\end{figure}

CF88 suggested an S factor at 300 keV of $S_{\rm 300}$ = 100 keV b.
This value is revised upward to 140 and 146 keV b by refs. 
\cite{2017RvMP...89c5007D} and \cite{1996ApJ...468L.127B},
respectively.
A more recent, higher-precision measurement \cite{2020PhRvL.124p2701S}
indicates an additional increase of about 10\% - 13\%, consistent with 
the value of 165 keV b reported in ref. \cite{2002ApJ...567..643K}.
These rates are compared in Figure \ref{fig:c12a_rate}.
HW02 used the rate recommended in ref. \cite{1996ApJ...468L.127B}.
TYU18 adopted a value equal to 1.2 times the CF88 rate,
which is lower than the widely used value ($\simeq$ 170 keV b; e.g., refs.
\cite{1993PhR...227...65W, 2007PhR...442..269W}).
We adopt the rate from ref. \cite{2002ApJ...567..643K}.

Considering an uncertainty of $\pm$ 1$\sigma$, we can cover the
effect of other rates. The $\sigma$ here is defined by the following formula:
\begin{equation}
e^{\sigma}=\sqrt{\frac{x_{\rm high}}{x_{\rm low}}}
\end{equation}
where $x_{\rm high}$ and $x_{\rm low}$ represent the high and low rates of the
reactions, respectively. Here, we use $\sigma_{C12\alpha}$
to represent the uncertainty of the $^{12}$C$(\alpha, \gamma)^{16}$O rate.
Because some previous studies still use the CF88 rate,
we use $f_{\rm CF88}$ to represent the multipliers on the CF88 rate.
Note, $\sigma_{C12\alpha}$ = $-$1, 0, and 1 in this study approximately
equals $f_{\rm CF88}$ = 1.34, 1.87, and 2.60, respectively.

\subsection{Treatment of Convection During Explosive Burning} \label{sec:TDC}

From the onset of O burning, the nuclear timescale decreases rapidly and becomes comparable
to the dynamical timescale, rendering the subsequent evolution inherently hydrodynamic.
The explosion then unfolds on a timescale of tens of seconds. Consequently, 
the convective timescale becomes comparable to or shorter than the convective turnover timescale.
Under these conditions, standard MLT is no longer physically reliable and
can even induce numerical instabilities. One practical approach is 
to turn-off convection entirely during the explosion phase
\cite{2019ApJ...887...72L, 2014AA...566A.146K}.
However, to capture the energy transport accurately under these conditions,
a time-dependent convection model is required \cite{2019ApJ...882...36M, Farmer_2019, 2023RAA....23a5014X}.
For a comprehensive discussion, we refer readers to
refs. \cite{2019ApJ...882...36M, 2020MNRAS.493.4333R}. Here, we provide only a brief overview.
In this framework, the average convective velocity $v_c$ is treated as an independent variable
governed by
\begin{equation} \label{equ:v_c}
\frac{\partial v_c}{\partial t} = \frac{v_{\rm MLT}^2-v_c^2}{\lambda}
\end{equation}
where $v_{\rm MLT}$ is the convective velocity predicted by MLT in stead-state and
$\lambda=\alpha_{\rm MLT} H_p$ is the mixing length.
For the timescale of convection much longer than a convective turnover timescale,
the convective velocities asymptotically approach $v_{\rm MLT}$, recovering the standard MLT.
Conversely, in regions where $v_{\rm MLT} = 0$,
Equation \ref{equ:v_c} yields convective decay on a timescale $\tau=\lambda/v_c$,
which becomes infinitely large as convective velocities are reduced.

The convergence and sensitivity have been well-tested in
previous studies by \cite{Farmer_2019, 2019ApJ...882...36M, 2020MNRAS.493.4333R, 
2022ApJ...937..112F, 2023RAA....23a5014X, 2026RAA....26g5011X}.
Ref. \cite{2020MNRAS.493.4333R} has investigated the differences in convective behavior
between this TDC model and the standard MLT approach.
They demonstrated that the thermonuclear oxygen burning remains purely radiative
until the outgoing pulse wave induces post-burning convection with the standard MLT model.
Conversely, the TDC model allows convective mixing to develop dynamically
during the active explosive oxygen burning phase.

\subsection{H-rich Model} \label{sec:H-rich}

We calculate a series of H-rich star models with ZAMS masses ranging from 
120 to 310 M$_\odot$ and terminate the evolution at the end of core He burning
($X(\rm He)< 10^{-4}$).
We adopt a mass interval of 20 M$_\odot$, and it is set finer near the boundaries
of PPISN/PISN and PISN/CCSN. The initial composition is set according to
Big Bang nucleosynthesis predictions \cite{2007ARNPS..57..463S}.
The mass fractions for $^{1}$H, $^{2}$H, $^{3}$He, $^{4}$He, and $^{7}$Li
are 0.7516, 4.01$\times 10^{-5}$, 2.39$\times 10^{-5}$, 0.2483, and 2.26$\times 10^{-9}$,
respectively, and no heavier nuclides than $A>7$ are produced in astrophysically
interesting abundances.

\begin{figure}[H]
\centering
\begin{minipage}[c]{0.48\textwidth}
\includegraphics [width=82mm]{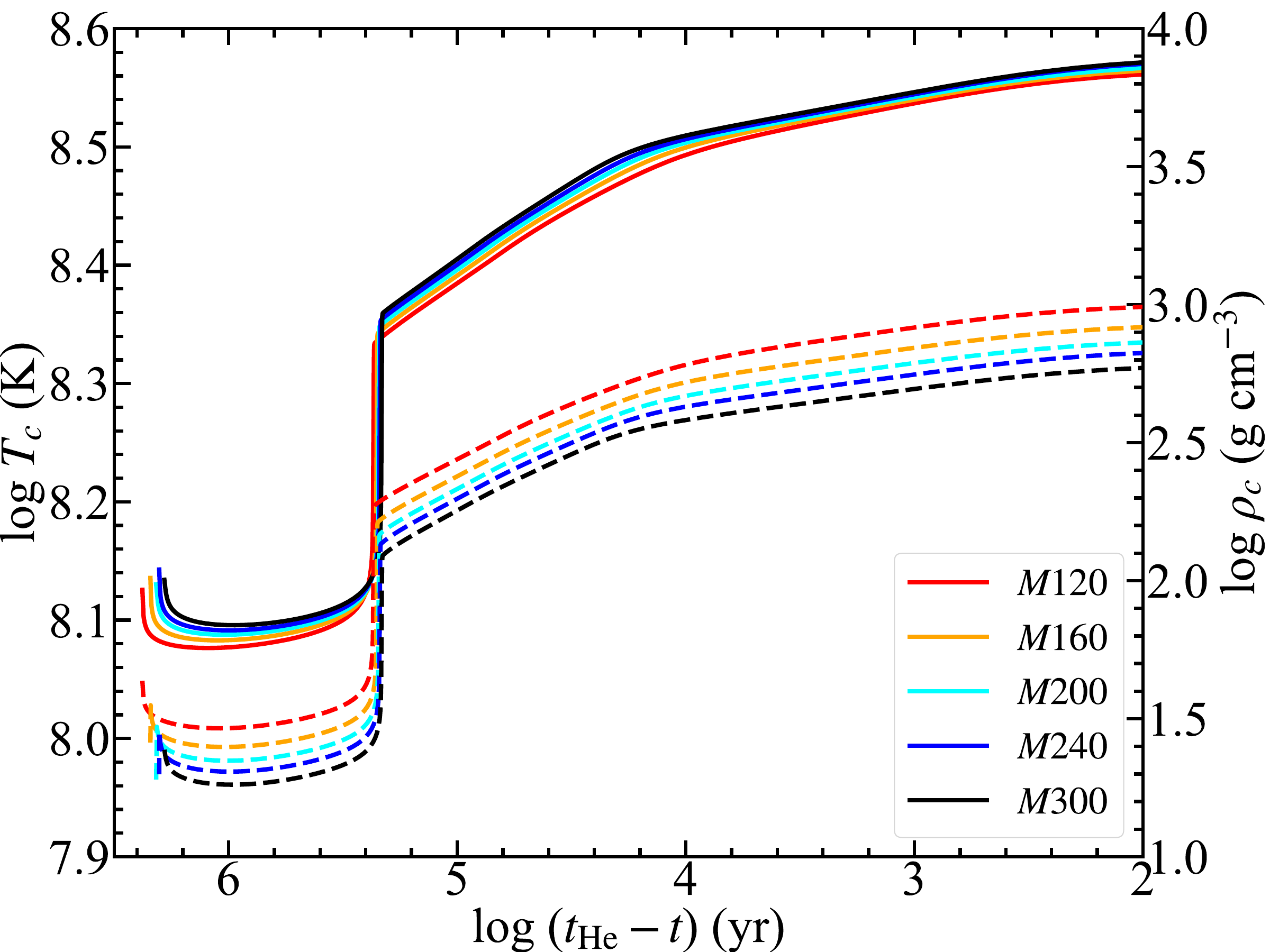}
\end{minipage}%
\caption{The central temperature (solid) and density (dashed) evolution
from ZAMS to the end of the core He burning for different initial masses.
$t_{\rm He}$ represents the time at the end of core He burning.
\label{fig:trho_evo}}
\end{figure}

Figure \ref{fig:trho_evo} shows that the massive metal-free stars start
the main sequence when the central temperature exceeds 10$^8$ K. 
They may not obtain sufficient energy from the pp-chain.
Instead, some C is produced via the 3$\alpha$ reaction,
and then H is burnt via the CNO cycle \cite{2001ApJ...550..890B}.
An initial neutron excess originates from the reaction chain of
$^{14}$N($\alpha$, $\gamma$)$^{18}$F(e$^+ \nu$)$^{18}$O.
The neutron excess, $\eta = \sum_i (N_i-Z_i)Y_i$, 
decreases from $1.6\times10^{-7}$ to $1.34\times10^{-7}$
at the end of core H burning for $M ({\rm ZAMS})$ = 120 M$_\odot$ to 310 M$_\odot$.

\begin{figure}[H]
\centering
\begin{minipage}[c]{0.48\textwidth}
\includegraphics [width=82mm]{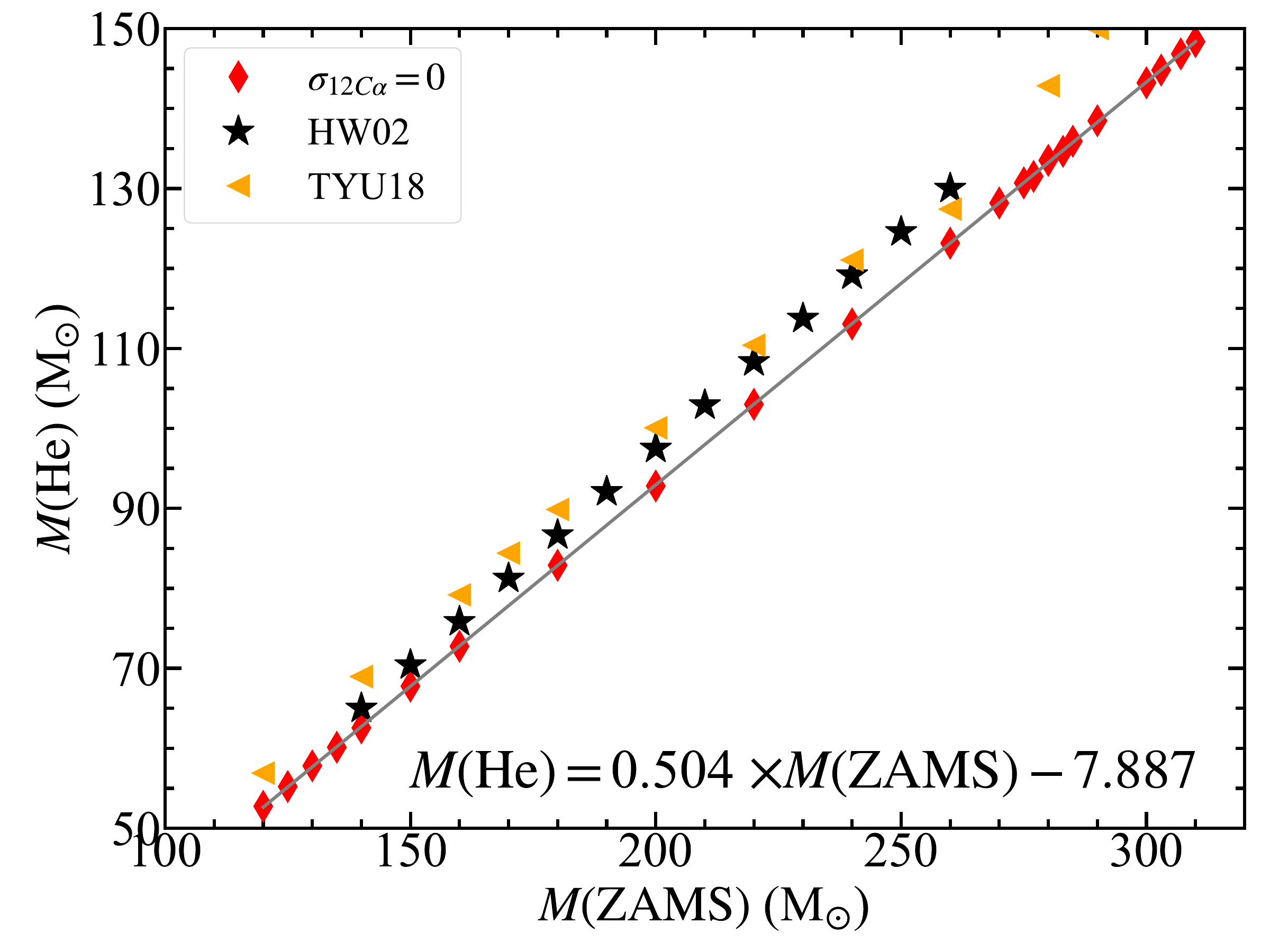}
\end{minipage}%
\caption{The He core mass as a function of $M ({\rm ZAMS})$.
The red diamonds, black stars, and orange triangles represent
$M$(He) in this study, HW02, and TYU18, respectively.
The gray straight line is obtained by fitting the red diamonds.
\label{fig:he_zams}}
\end{figure}

We defined the boundary of He core mass, $M(\rm He)$, at $M_r$
where $X$(H) decreases down to $10^{-4}$.
The He core masses as a function of $M ({\rm ZAMS})$s are
shown in Figure \ref{fig:he_zams}. We obtain the green straight line of
\begin{equation} \label{equ:he_zams}
M({\rm He}) = 0.504\times M(\rm ZAMS) - 7.887\, {\rm M}_\odot
\end{equation}
by fitting the red diamonds.
Because $^{12}$C$(\alpha, \gamma)^{16}$O is very weak during shell H burning,
this reaction rate have less impact on $M(\rm He)$. 
We just use $\sigma_{C12\alpha}=0$ in H-rich models and compare the
$M(\rm ZMAS)$ - $M(\rm He)$ relation in this study with HW02 and TYU18.
The $M(\rm He)$s in this study are smaller than those reported in the
other two studies by $\sim8$\%, because both these two studies adopted
a larger overshooting mixing.

\subsection{He star Model} \label{sec:He_star}

A series of almost pure He-star models is then calculated from
the zero-age horizontal branch (ZAHB) and through the following stages: 
(1) The pulsational pair instabilities (PPIs) occur during explosive O
burning and then the Fe-core collapse (PPISN),
(2) Explode as PISNe,
(3) A Fe core is formed and then collapses without PPI (Fe CC).

\begin{table}[H]
\caption{The composition at the He ignition for $M ({\rm ZAMS})$ = 240 M$_\odot$}
\label{tab:composition}
\centering
\begin{tabular}{cccc}
\toprule
Isotopes & Mass fraction & Isotopes & Mass Fraction \\
\midrule
$^{4}$He      & 0.99998       & $^{18}$F      & 5.43$\times 10^{-12}$      \\
$^{12}$C      & 1.68 $\times 10^{-05}$      & $^{19}$F      & 5.72$\times 10^{-11}$      \\
$^{13}$C      & 2.71 $\times 10^{-12}$      & $^{20}$Ne     & 1.35$\times 10^{-10}$      \\
$^{14}$N      & 9.01 $\times 10^{-07}$      & $^{21}$Ne     & 2.60$\times 10^{-11}$      \\
$^{15}$N      & 7.32 $\times 10^{-09}$      & $^{22}$Ne     & 1.21$\times 10^{-09}$      \\
$^{16}$O      & 8.01 $\times 10^{-07}$      & $^{28}$Si     & 5.97$\times 10^{-13}$      \\
$^{17}$O      & 2.39 $\times 10^{-08}$      & $^{29}$Si     & 4.94$\times 10^{-14}$      \\
$^{18}$O      & 1.07 $\times 10^{-07}$      & $^{30}$Si     & 1.85$\times 10^{-12}$      \\
\bottomrule
\end{tabular}
\end{table}

We construct almost pure He-star models from H-rich progenitors calculated in
Section \ref{sec:H-rich} following the method from HW02 \cite{2002ApJ...567..532H}.
The almost pure He-star models are mapped with identical chemical
compositions and entropy profiles from H-rich models at ZAHB,
when the luminosity of He burning exceeds 90\% of the total luminosity.
The initial masses of He-star models keep identical with the He core
masses at the end of core He burning to account for their growth due to
the shell H-burning.
Different from HW02, which adopts a universal composition from a single
representative model, we calculate a dedicated H-rich model for each
He-star model to capture mass-dependent variations in entropy and isotopic profiles.
We denote H-rich model with $M ({\rm ZAMS})$ = 240 M$_\odot$ as ``M240''
and the corresponding He-star model with $M(\rm He)$ = 113 M$_\odot$ as
``He113''. The main isotopic composition of M240 at ZAHB is shown in
Table \ref{tab:composition} as an example. 

\begin{table*}[]
\caption{The comparison of the core properties at the end of He burning between
the H-rich star models with $M(\rm ZAMS)$ = 240 M$_\odot$ and the He-star models
with $M(\rm He)$ = 113 M$_\odot$. The boundary of the CO core is defined at $M_r$
where $X$(He) decreases down to 10$^{-4}$. $X$(C)/$X$(O) represents the ratio
of the mass fraction of $^{12}$C to $^{16}$O in the center. $\eta$ is neutron excess.}
\label{tab:He-H}
\centering
\begin{tabular}{cccccccc}
\toprule
Model & $M(\rm He)$ & $M(\rm CO)$ & log $T_c$ & log $\rho_c$ & $X$($^4$He) & $X$(C)/$X$(O) & $\eta$\\
      & (M$_\odot$) & (M$_\odot$) &    (K)    & (g cm$^{-3}$)&      \\
\midrule
H-rich   & 113.01 & 100.46 & 8.576 & 2.843 & 7.1$\times$10$^{-5}$ & 0.055 & 1.437$\times$10$^{-7}$ \\
He-star  & 113.00 & 101.38 & 8.580 & 2.891 & 6.9$\times$10$^{-5}$ & 0.063 & 1.440$\times$10$^{-7}$\\
\bottomrule
\end{tabular}
\end{table*}

\begin{figure}[H]
\centering
\begin{minipage}[c]{0.48\textwidth}
\includegraphics [width=82mm]{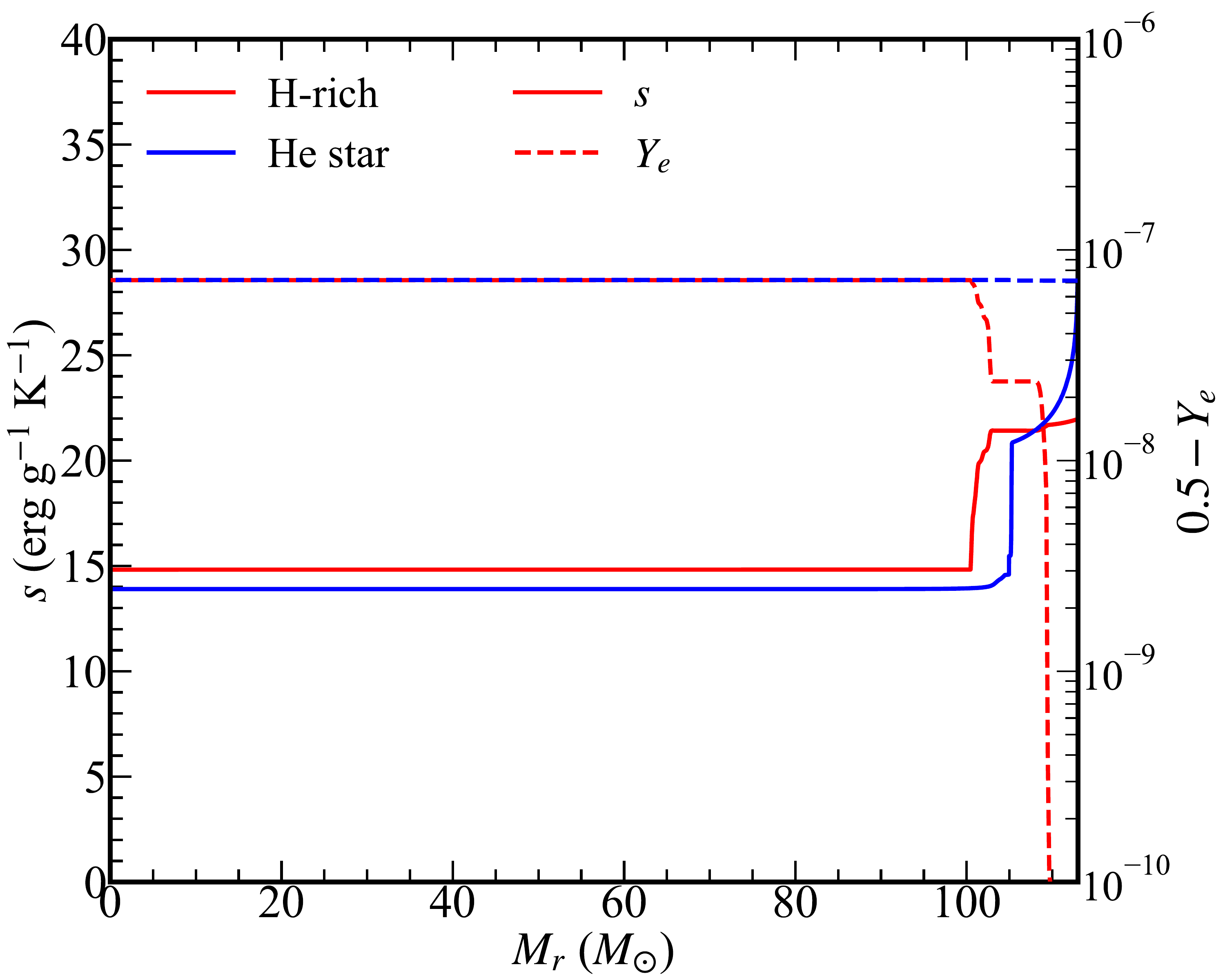}
\end{minipage}%
\caption{The mass distribution of entropy and $Y_{\rm e}$ for
He stars with $M(\rm He)=$ 113 M$_\odot$ and H-rich stars with
$M(\rm ZAMS)=$ 240 M$_\odot$, respectively.
\label{fig:s_ye}}
\end{figure}

To demonstrate that the He-star models faithfully preserve the core structures
of their H-rich progenitors, we compare the entropy and $Y_{\rm e}$ profiles
(Figure \ref{fig:s_ye}) and other key core properties (Table \ref{tab:He-H})
between the M240 model and its corresponding He113 model
at the end of core He burning.
In Figure \ref{fig:s_ye}, the discrepancies observed in the He shell
between the two models are an expected consequence of removing the hydrogen envelope.
The M240 model exhibits more physically realistic abundance gradients
and entropy variations in the He-rich shell and H/He boundary.
In contrast, the surface entropy of the He113 model increases rapidly
because an additional pressure is applied to ensure numerical stability
during the He-burning evolution \cite{1974PASJ...26..129N}. However, 
these discrepancies do not affect the structure of the CO core significantly.
Inside the CO core ($M_r \leq$ 100 M$_\odot$), both models present
excellent agreement in $Y_{\rm e}$ profiles,
and core entropy in the M240 model is slightly higher than that in the He113 model.
In Table \ref{tab:He-H}, the CO core mass, temperature, density,
and neutron excess differ by $<3\%$ between the two models.

The (P)PIs process is hydrodynamic and generates significant shocks.
To simulate these shocks, the hydrodynamic module is required
\cite{2019ApJ...887...72L}.
The Harten–Lax–van Leer–contact (HLLC) solver method \cite{1994ShWav...4...25T}
provides improved shock capturing and energy conservation. 
Because of computational reasons, the HLLC solver is only employed once
a star departs from hydrostatic equilibrium.
Once the shocks reach the surface of the star, the materials whose velocity
exceeds the escape velocity will be removed \cite{2016MNRAS.457..351Y}.
A new stellar structure is then generated with identical mass, chemical
composition and entropy to the prior configuration sans ejecta.
Thereafter, calculations switch to a hydrostatic solver to continue the
evolutionary process. For PISNe, the calculation will stop until all 
materials become unbound.
To avoid the PPIs being interrupted by the code stop control,
we set a large value for the limit of the inward velocity of the
Fe core (fe\_core\_infall\_limit=1d99).
The Fe CC would cease when the central temperature exceeds 10$^{10}$ K.
All the PISNe models are calculated until the entire star expands above
the escape velocity, where the central temperature is usually lower than
10$^{8}$ K.

\section{The evolution and nucleosynthesis of Pop III PISNe} \label{sec:popIII}

Given that the evolution of Pop III PISNe has been extensively detailed
by HW02 and TYU18, we do not reiterate those discussions here.
Instead, Section \ref{sec:evo_exp} highlights the discrepancies between
our key results and those of earlier studies,
specifically assessing the sensitivity of these outcomes to the
$^{12}$C$(\alpha, \gamma)^{16}$O reaction rate. 
In Section \ref{sec:oo}, we conduct a sensitivity analysis of the
$^{16}$O+$^{16}$O reaction rate.
In Section \ref{sec:mass_dep}, we present predicted elemental abundances
as a function of $M(\rm He)$ and quantify the theoretical uncertainties
arising from variations in the $^{12}$C$(\alpha, \gamma)^{16}$O and
$^{16}$O+$^{16}$O reaction rates. Our results are compared with previous
studies and observations in Section \ref{sec:observe}.

\subsection{The effect of $^{12}$C$(\alpha, \gamma)^{16}$O reaction rate on the 
Evolution of Pop III PISNe}  \label{sec:evo_exp}

The advanced burning phases of massive stars,
from the central He exhaustion to the onset of the explosion/collapse,
are primarily governed by both the CO core mass and the mass fraction of $^{12}$C,
$X$($^{12}$C), at He exhaustion \cite{2001ApJ...558..903I,
2020ApJ...890...43C, 2025arXiv250211012X}.

\begin{figure}[H]
\centering
\begin{minipage}[c]{0.48\textwidth}
\includegraphics [width=80mm]{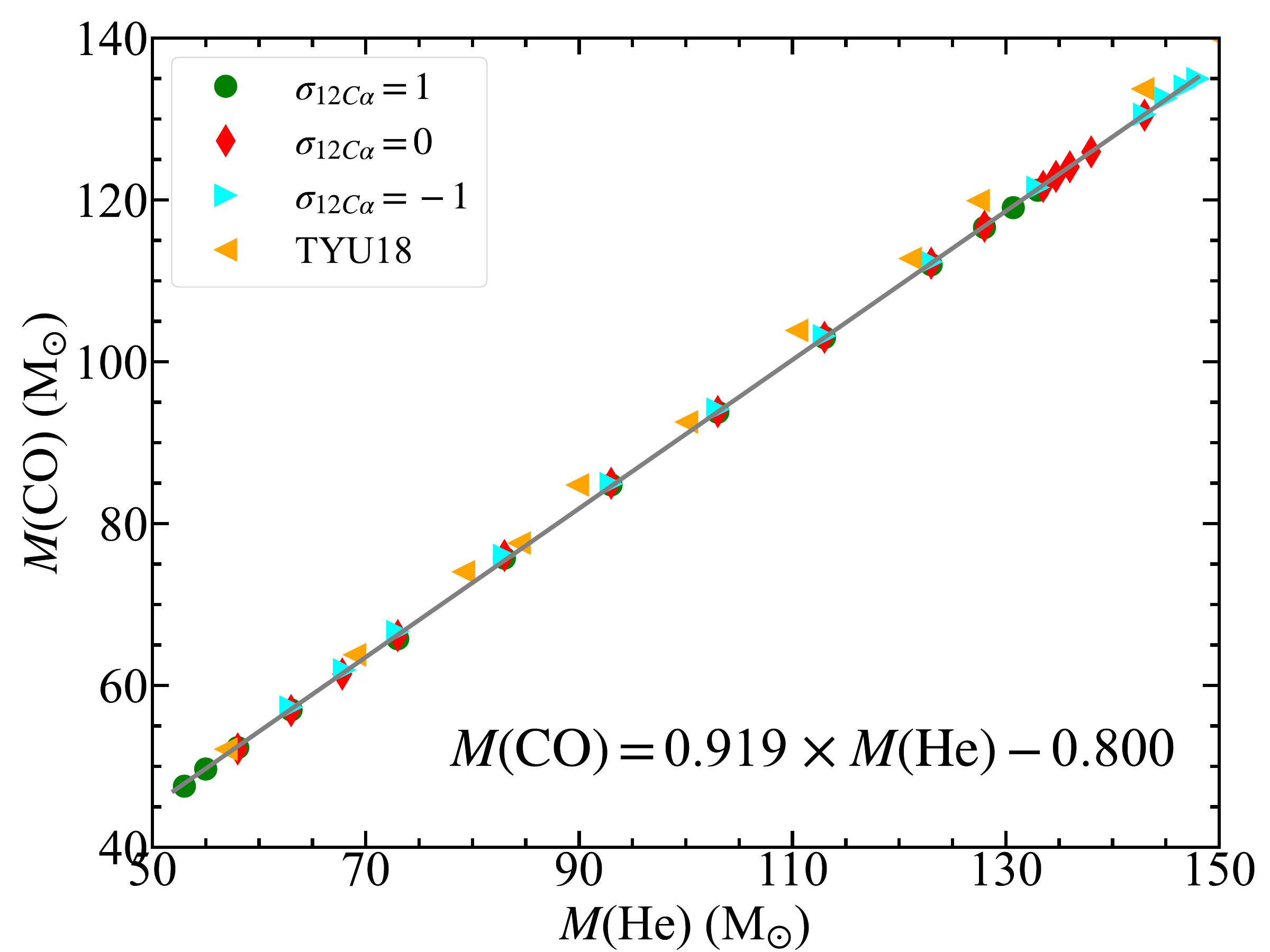}
\end{minipage}%
\caption{The CO core mass as a function of $M ({\rm He})$ for
$\sigma_{C12\alpha}=0, \, \pm1$.
The orange triangles represent $M(\rm He)$ reported in TYU18.
The gray straight line is obtained by fitting the data in this study.
\label{fig:mco_sigma}}
\end{figure}

The relation between $M(\rm He)$ and $M(\rm CO)$
is shown in Figure \ref{fig:mco_sigma}. 
All models follow a tight linear relation:
\begin{equation} \label{equ:he_co}
M({\rm CO}) = 0.919\times M(\rm ZAMS) - 0.800\, {\rm M}_\odot
\end{equation}
indicating that $\sigma_{C12\alpha}$ has a negligible effect on $M(\rm CO)$.
Our $M(\rm CO)$ agrees well with those reported in TYU18.

\begin{figure}[H]
\centering
\begin{minipage}[c]{0.48\textwidth}
\includegraphics [width=80mm]{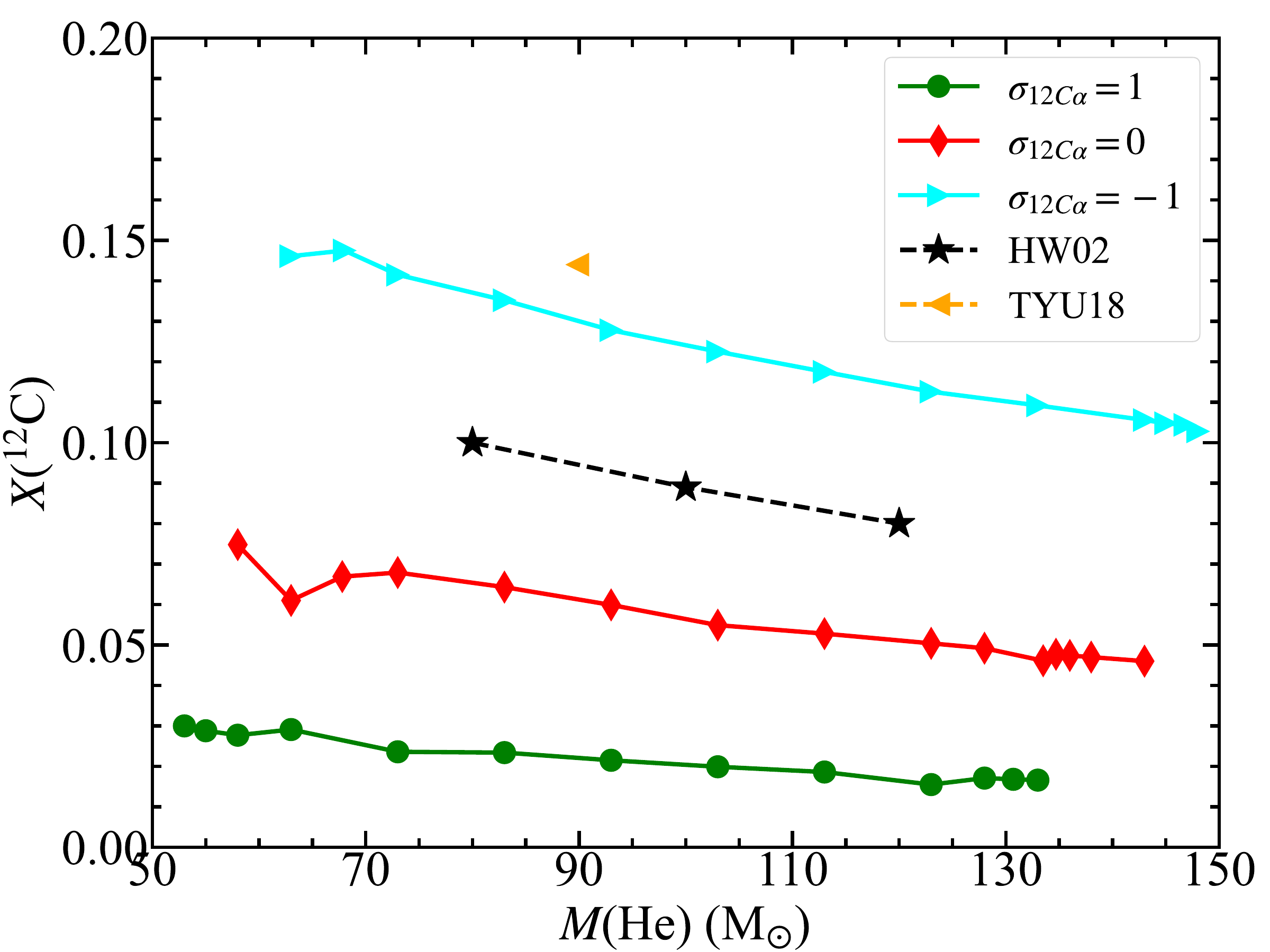}
\end{minipage}%
\caption{The mass fraction of $^{12}$C left in the core after He exhaustion
as a function of $M ({\rm He})$ for $\sigma_{C12\alpha} = 0,\, \pm1$.
The black stars and orange triangles represent those reported in HW02 and TYU18.
\label{fig:xc_sigma}}
\end{figure}

As shown in Figure~\ref{fig:xc_sigma},
$X$($^{12}$C) decreases with increasing $M(\rm He)$ or $\sigma_{C12\alpha}$,
with the latter having a substantially larger impact.
Previous studies (HW02; TYU18) generally predicted higher $X$($^{12}$C) at the He exhaustion
than our fiducial models, primarily due to their adoption of the different
$^{12}$C$(\alpha, \gamma)^{16}$O and $3\alpha$ reaction rate and convection treatments.

\begin{figure*}[t]
\centering
\begin{minipage}[c]{0.8\textwidth}
\includegraphics [width=140mm]{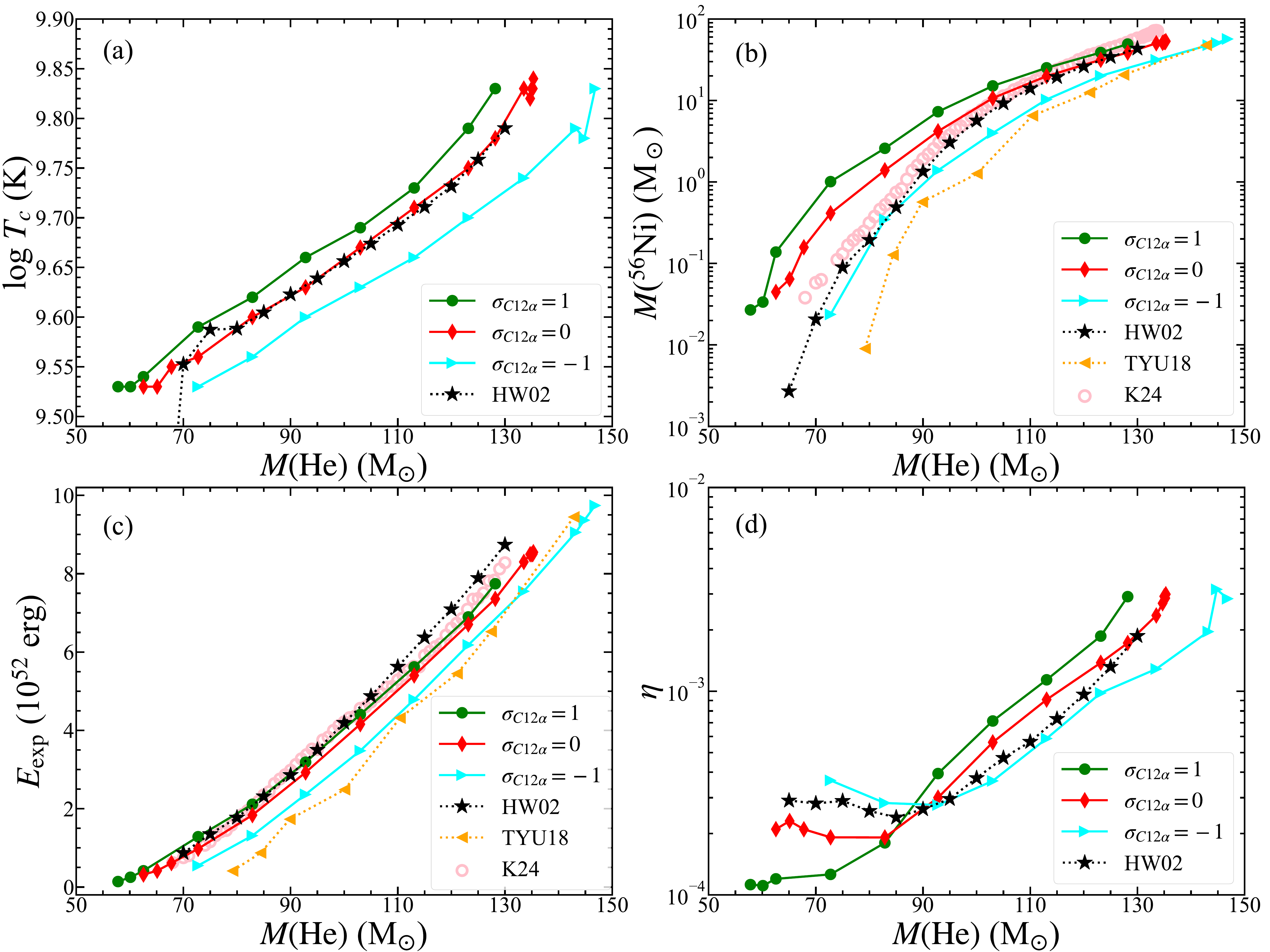}
\end{minipage}%
\caption{The maximum central temperature $T_{\rm c}$ (a), total mass of
$^{56}$Ni (b), explosion energy $E_{\rm exp}$ (c) and neutron excess $\eta$
(d) as a function of He core mass $M(\rm He)$ for
$\sigma_{C12\alpha} = 0, \, \pm1$.
The black stars, orange triangles, and pink circles show the results
from HW02 \cite{2002ApJ...567..532H}, TYU18 \cite{Takahashi_2018_1},
and K24 \cite{2024MNRAS.531.2786K}.
Note that $E_{\rm exp}$ in HW02 is the kinetic energy evaluated at carbon ignition,
whereas this work, TYU18, and K24 evaluate $E_{\rm exp}$ using the
total energy after the explosion.  
\label{fig:exp_mhe}}
\end{figure*}

Figure~\ref{fig:exp_mhe} compares the maximum central temperature
($T_{\rm c}$), $^{56}$Ni mass ($M$($^{56}$Ni)), explosion energy ($E_{\rm exp}$),
and neutron excess ($\eta$) for different $\sigma_{C12\alpha}$
with HW02, TYU18, and K24 \cite{2024MNRAS.531.2786K}.
Here, $M$($^{56}$Ni), $E_{\rm exp}$, and $\eta$ are defined
when all the materials become unbound.
$E_{\rm exp}$ is defined as a sum of the gravitational, kinetic, and internal energies.
Ultimately, $E_{\rm exp}$ is dominated by the kinetic energy.

We find that $T_{\rm c}$, $M$($^{56}$Ni), and $E_{\rm exp}$ are higher
for higher $M(\rm He)$ and $\sigma_{C12\alpha}$.
This trend is directly linked to the strength of shell carbon burning:
higher $M(\rm He)$ and $\sigma_{C12\alpha}$ leave less $^{12}$C fuel at the He
exhaustion (Figure \ref{fig:xc_sigma}), and thus weaken the heating effect of shell C-burning.
Consequently, the CO cores contract further to compensate for the neutrino losses,
and reach higher $T_c$, thus producing more energetic explosions with larger $^{56}$Ni yields.
The neutron excess $\eta$ remains nearly constant for $M(\rm He)<$ 83 M$_\odot$,
since these models explode before the Si ignition.
A significant $\eta$ enhancement occurs only during Si burning and subsequent NSE processes,
which will be discussed in Section \ref{sec:nuc_PopII}.

The results of HW02 fall between our $\sigma_{C12\alpha} = -1$ and 0 models,
while TYU18 predicts lower $M$($^{56}$Ni) and $E_{\rm exp}$ than even our $\sigma_{C12\alpha} = -2$ case.
These systematic offsets align with the $X$($^{12}$C) trends in
Figure~\ref{fig:xc_sigma}. This indicates that discrepancies with previous studies
are primarily attributed to the different $^{12}$C$(\alpha, \gamma)^{16}$O
reaction rates. The apparently higher $E_{\rm exp}$ in HW02 than our
$\sigma_{C12\alpha} = 1$ model arises
because they define it as the kinetic energy at the C ignition.
We also compare our $E_{\rm exp}$ and $M$($^{56}$Ni) with K24,
who adopted a very fine mass grid ($\Delta M = 0.01\, M_\odot$) at $Z = 10^{-5}$.
Their $^{12}$C$(\alpha, \gamma)^{16}$O rate, adopted from STARLIB \cite{2013ApJS..207...18S},
is slightly higher than ours.

\subsection{The effect of $^{16}${\rm O+}$^{16}${\rm O} Reaction Rate} \label{sec:oo}

The $^{16}$O+$^{16}$O reaction rate plays a pivotal role in the PISN explosions.
However, current models generally rely on the outdated prescription from the CF88 compilation,
which lacks temperature-dependent uncertainties. 
Due to experimental challenges associated with oxygen targets and
the existence of complex decay channels, 
all existing measurements are restricted to energies above $E_{\rm c.m.} =$ 6.7 MeV,
slightly higher than the median of the Gamow window
\cite{2025JPhG...52k5103I, 2025AA...702A..86D}.
Furthermore, as previous measurements have focused primarily on S-factors,
no reaction rate incorporating temperature-dependent uncertainties is currently
available. While a global scaling factor may not fully capture
the temperature-dependent physics,
it serves as a necessary approximation consistent with previous studies
\cite{2018ApJS..234...19F, Farmer_2020}.
Therefore, we conduct a sensitivity experiment using the $M(\rm He)$ = 113 $M_{\odot}$ model,
where the CF88 rate is scaled by a fixed factor of 10 across the entire temperature 
range to bracket the potential impact of this uncertainty.

\begin{figure}[H]
\centering
\begin{minipage}[c]{0.48\textwidth}
\includegraphics [width=82mm]{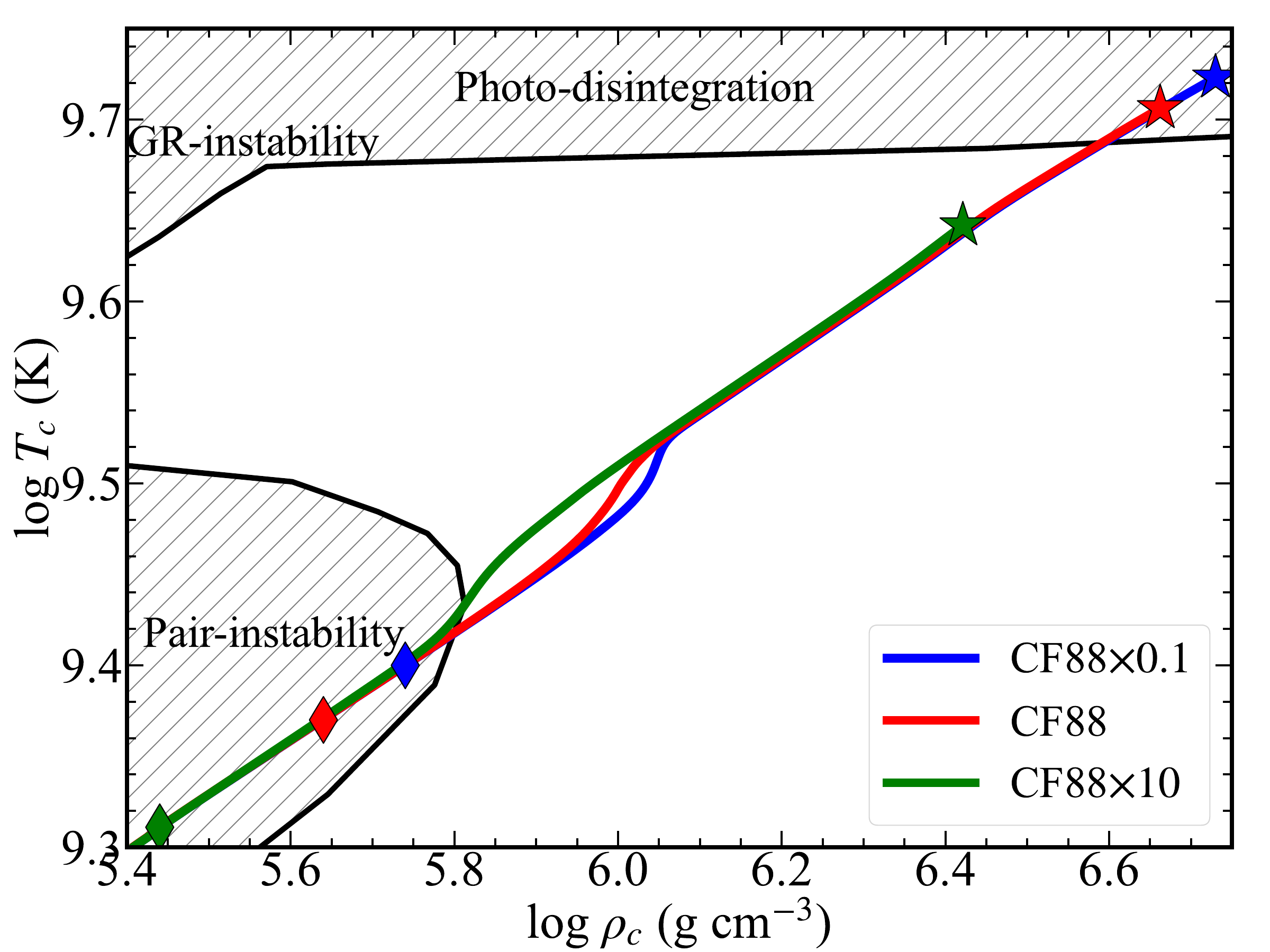}
\end{minipage}%
\caption{The central temperature against the central density for the evolution of
the star with $M(\rm He)$ = 113 M$_{\odot}$ for $f_{\rm 16O}$ = 0.1 (blue), 1 (red),
10 (green), respectively.
The ignition points and maximum temperature points are indicated
by ``diamonds" and ``stars".
In the hatched region, stars are dynamically unstable due to the
electron-positron pair creation (indicated as ``Pair-instability Region''),
general relativistic effects (``GR instability'')
and the photo-disintegration of matter in nuclear statistical equilibrium (NSE)
at $Y_{\rm e}=$ 0.5 (``photo-disintegration'').
\label{fig:pisn_oo_rate}}
\end{figure}

In Figure \ref{fig:pisn_oo_rate}, the diagram of $T_c$ against $\rho_c$ for
the evolution of the star with $M(\rm He)$ = 113 M$_{\odot}$ is shown.
Lower $f_{\rm 16O}$ requires higher temperatures to ignite central O burning,
delaying ignition and driving further contraction via neutrino losses,
which ultimately leads to a higher peak $T_c$ during the explosion.
Notably, models with $f_{\rm 16O}$ may explode before entering the
photo-disintegration regime,
resulting in suppressed Fe yields due to the lower explosion temperature.

\subsection{The Mass Dependence and Reaction Rate Dependence of Nucleosynthesis Yields}
\label{sec:mass_dep}

When the hydrodynamic calculation is finished, a further decay process is
considered by calculating an additional 10$^{10}$ yr with a constant
temperature of 10$^{3}$ K and the density of 10$^{-10}$ g cm$^{-3}$.

\begin{figure}[H]
\centering
\begin{minipage}[c]{0.48\textwidth}
\includegraphics [width=85mm]{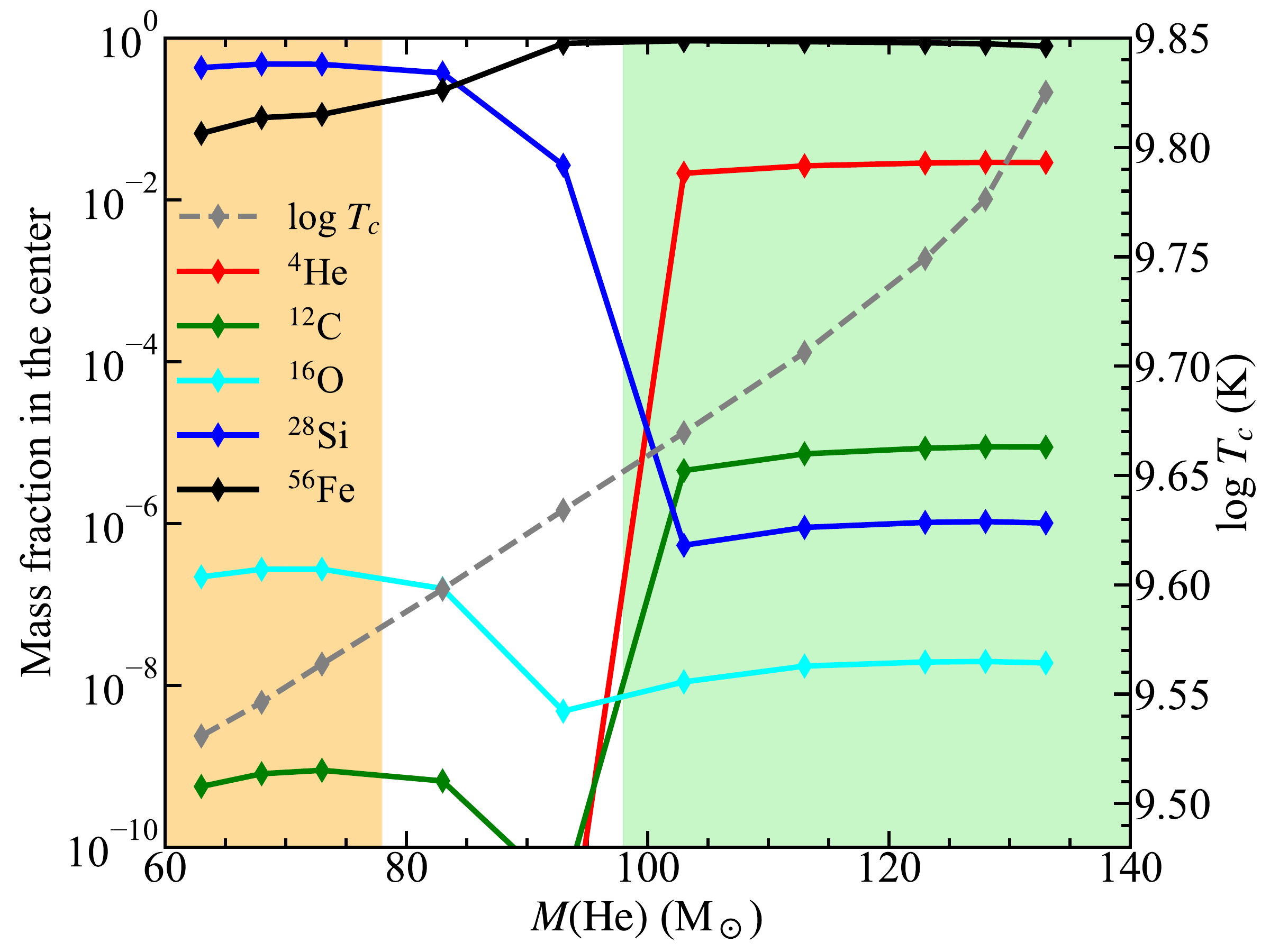}
\end{minipage}%
\caption{The central mass fraction of $^{4}$He, $^{12}$C, $^{16}$O, $^{28}$Si,
$^{56}$Fe after explosion and the maximum of log $T_{\rm c}$ as a function
of $M(\rm He)$. The orange, write, and light green regions represent that
the stars reach O burning, incomplete Si burning, and the NSE process
during the explosion.
\label{fig:ctrx}}
\end{figure}  

Figure \ref{fig:ctrx} summarizes the changes in the central compositions
after the explosions for different $M(\rm He)$.
We find that He63, He68, and He73 models explode before Si burning ignition
with central $X$(Si) = 0.432, 0.478, 0.475, and $X$(Fe) = 0.07, 0.10, and 0.11, respectively.
He83 and He93 models undergo incomplete Si burning during the explosion,
with a central $X$(Si) = 0.37 and 0.03, and $X$(Fe) = 0.23 and 0.86.
Only those with $M(\rm He)>$ 93 M$_\odot$, corresponding
to $M(\rm ZAMS)>$ 200 M$_\odot$, can undergo the NSE process.
The iron mass, in fact, $M$($^{56}$Ni), for different models has been
shown in Figure \ref{fig:exp_mhe} (b) and Table \ref{tab:my-table}.

To facilitate comparison with astronomical observations.
We show the abundance ratios of element i relative to element j.
The abundance ratio is defined as:
\begin{equation} \label{equ:x_fe}
[\frac{N_i}{N_j}] \equiv {\rm log}(\frac{N_i}{N_j})- {\rm log}((\frac{N_i}{N_j})_\odot)
\end{equation}
where $N_i$ and $N_j$ represent the number of elements i and j, respectively.
The solar abundance adopted here is from Ref. \cite{2009LanB...4B..712L}. 

\begin{table*}[htbp]
\caption{Model Properties. $M(\rm ZAMS)$, $M(\rm He)$, and $M(\rm CO)$ represent the
initial mass, He core mass, and CO core mass, respectively.
$X$($^{12}$C) is the central mass fraction of $^{12}$C at the end of core He burning.
Fate is only PISN because the models beyond PISN are not listed in this table.
$E_{\rm tot}$ and $M$($^{56}$Ni) are the total energy and ejected $^{56}$Ni mass.
log $\rho_{\rm max}$, log $T_{\rm max}$, and $\eta$ are the maximum central density,
temperature, and neutron excess during the explosion.}
\label{tab:my-table}
{\normalsize
\begin{tabular}{cccccccccc}
\toprule
$M(\rm ZAMS)$   & $M(\rm He)$  & $M(\rm CO)$  & $X$($^{12}$C)  & Fate  & $E_{\rm tot}$     & $M$($^{56}$Ni)     & log $\rho_{\rm max}$  & log $T_{\rm max}$  & $\eta$ \\
($M_{\odot}$)     & ($M_{\odot}$)  & ($M_{\odot}$)  &         &      & (10$^{52}$ erg) & ($M_{\odot}$) & (g cm$^{-3}$)  & (K)         \\
\midrule
\multicolumn{10}{c}{$\sigma=1$}      \\
130             &   57.80      &    52.31     &  0.0277 &  PISN &   0.14   &   0.03  &   6.23   &   9.53  & 1.12$\times10^{-4}$ \\
140             &   62.55      &    56.97     &  0.0291 &  PISN &   0.42   &   0.14  &   6.27   &   9.54  & 1.20$\times10^{-4}$ \\
160             &   72.75      &    65.77     &  0.0236 &  PISN &   1.29  &    1.01  &   6.38   &   9.59  & 1.26$\times10^{-4}$ \\
180             &   82.88      &    75.73     &  0.0234 &  PISN &   2.11  &    2.59  &   6.44   &   9.62  & 1.80$\times10^{-4}$ \\
200             &   92.80      &    84.79     &  0.0215 &  PISN &   3.19  &    7.29  &   6.53   &   9.66  & 3.94$\times10^{-4}$ \\
220             &   103.01     &    93.76     &  0.0199 &  PISN &   4.41  &    15.1  &   6.63   &   9.69  & 7.14$\times10^{-4}$ \\
240             &   113.06     &    102.98    &  0.0186 &  PISN &   5.62  &    25.2  &   6.76   &   9.73  & 1.14$\times10^{-3}$ \\
260             &   123.16     &    111.99    &  0.0155 &  PISN &   6.90  &    38.3  &   6.97   &   9.79  & 1.86$\times10^{-3}$ \\
270             &   128.19     &    116.59    &  0.0171 &  PISN &   7.74  &    49.3  &   7.15   &   9.83  & 2.91$\times10^{-3}$ \\
\midrule
\multicolumn{10}{c}{$\sigma=0$}      \\
140             &   62.55      &   56.89      &  0.0766 &  PISN &   0.32   &   0.04  &   6.21  &   9.52  & 2.10$\times10^{-4}$   \\
150             &   67.77      &   61.41      &  0.0669 &  PISN &   0.61   &   0.16  &   6.26  &   9.55  & 2.10$\times10^{-4}$   \\
160             &   72.75      &   66.15      &  0.0679 &  PISN &   0.98   &   0.41  &   6.30  &   9.56  & 1.91$\times10^{-4}$   \\
180             &   82.88      &   76.05      &  0.0643 &  PISN &   1.84   &   1.39  &   6.38  &   9.60  & 1.91$\times10^{-4}$   \\
200             &   92.80      &   84.98      &  0.0599 &  PISN &   2.93   &   4.19  &   6.45  &   9.63  & 3.00$\times10^{-4}$   \\
220             &   103.01     &   93.84      &  0.0549 &  PISN &   4.16   &   10.7  &   6.55  &   9.67  & 5.61$\times10^{-4}$   \\
240             &   113.06     &   103.02     &  0.0528 &  PISN &   5.40   &   19.9  &   6.66  &   9.71  & 9.10$\times10^{-4}$   \\
260             &   123.16     &   112.18     &  0.0504 &  PISN &   6.71   &   31.6  &   6.81  &   9.75  & 1.38$\times10^{-3}$   \\
270             &   128.19     &   116.74     &  0.0492 &  PISN &   7.36   &   38.7  &   6.91  &   9.78  & 1.72$\times10^{-3}$   \\
280             &   133.52     &   121.67     &  0.0461 &  PISN &   8.30   &   50.4  &   7.10  &   9.83  & 2.35$\times10^{-3}$   \\
\midrule
\multicolumn{10}{c}{$\sigma=-1$}       \\
160             &   72.75      &   66.64      &  0.1415 &  PISN &   0.55   &  0.02  &   6.20  &   9.53  & 3.65$\times10^{-4}$   \\
180             &   82.88      &   76.05      &  0.1352 &  PISN &   1.31  &   0.35  &   6.26  &   9.56  & 2.82$\times10^{-4}$   \\
200             &   92.80      &   84.92      &  0.1278 &  PISN &   2.37  &   1.39  &   6.35  &   9.60  & 2.76$\times10^{-4}$   \\
220             &   103.01     &   94.12      &  0.1225 &  PISN &   3.48  &   3.99  &   6.42  &   9.63  & 3.63$\times10^{-4}$   \\
240             &   113.06     &   103.18     &  0.1175 &  PISN &   4.79  &   10.3  &   6.50  &   9.66  & 5.86$\times10^{-4}$   \\
260             &   123.16     &   105.96     &  0.1126 &  PISN &   6.18  &   20.1  &   6.61  &   9.70  & 9.80$\times10^{-4}$   \\
280             &   133.52     &   121.53     &  0.1092 &  PISN &   7.55  &   31.6  &   6.74  &   9.74  & 1.29$\times10^{-3}$   \\
300             &   143.22     &   130.57     &  0.1056 &  PISN &   9.06  &   47.8  &   6.96  &   9.79  & 1.96$\times10^{-3}$   \\
\bottomrule
\end{tabular}
}
\end{table*}

\begin{figure*}[htbp]
\centering
\begin{minipage}[c]{0.9\textwidth}
\includegraphics [width=160mm]{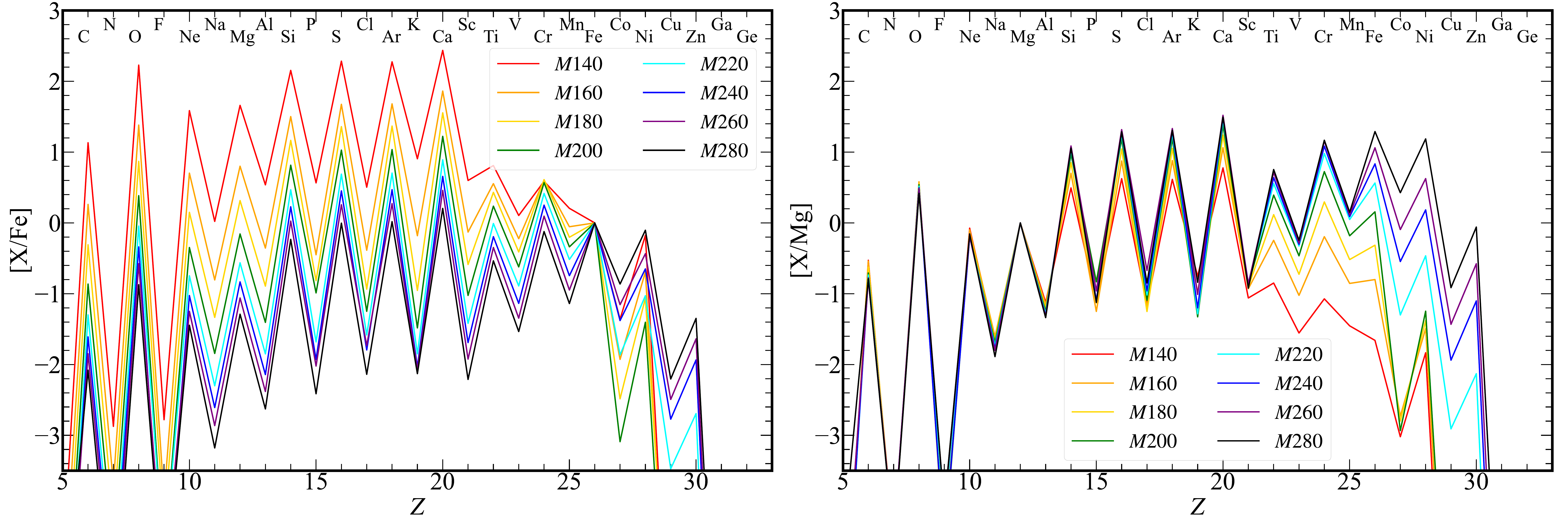}
\end{minipage}%
\caption{Abundance patterns of PISN yields normalized by iron yield (left)
and magnesium yield (right) for $\sigma_{C12\alpha}$ = 0. The models are marked by
the $M$(ZAMS) instead of $M$(He).
\label{fig:pisn_yield}}
\end{figure*}

Figure \ref{fig:pisn_yield} presents the abundance patterns for $\sigma_{C12\alpha}$ = 0.
The abundance patterns of $\sigma_{C12\alpha}$ = $\pm$1 are shown in
Figure \ref{fig:pisn_yield000}.
In the left panel, the abundance ratios relative to iron yield are presented. 
Most ratios span a wide range because the iron yield is sensitive to $M(\rm He)$,
as shown in Figure \ref{fig:exp_mhe} (b).
Pronounced odd-even effects are observed from [Ne/Fe] to [Sc/Fe]
due to the low neutron excess in the Pop III stars (see Figure \ref{fig:exp_mhe} (d)).
Lower central temperature at the bounce leads to lower abundances of Fe-peak
elements (e.g., [Cu/Fe]$<-2.1$ and [Zn/Fe]$<-1.3$), and essentially no s- or r-process
elements are produced. These features have been mentioned or discussed by many
studies \cite{2002ApJ...565..385U, 2002ApJ...567..532H, 2013ARAA..51..457N, 
Takahashi_2018_1, 2019MNRAS.487.4261S, 2023Natur.618..712X,
2024MNRAS.527.4790J, 2024ApJ...967L..22V}.

In the right panel, the abundance ratios are normalized to magnesium,
which remains nearly constant at about 1.47 -- 1.63 M$_{\odot}$ across all models,
indicating minimal dependence on $M(\rm He)$. The abundances of these 
elements become increasingly dispersed as $M(\rm He)$ increases.
Lighter elements (from C to Al) are predominantly produced in
the CO shell and are insensitive to explosive burning and thus,
regardless of variations in $M(\rm He)$.
Intermediate elements (from Si to Sc) are mainly synthesized in OSi shells and
increase with $M(\rm He)$ increasing.
The yields of Fe-peak elements are highly sensitive to the maximum temperature during
the explosion.

\begin{figure}[H]
\centering
\begin{minipage}[c]{0.48\textwidth}
\includegraphics [width=85mm]{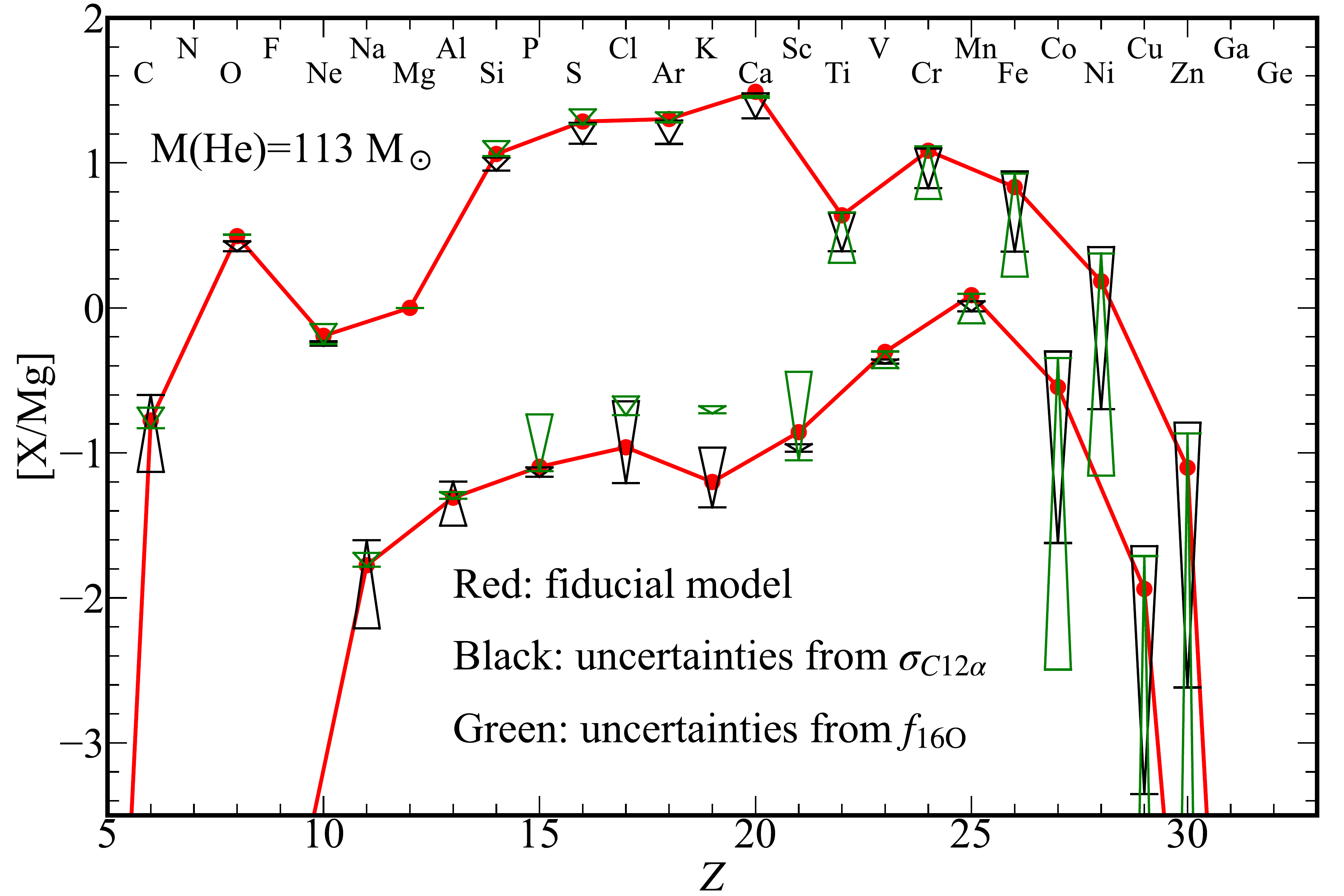}
\end{minipage}%
\caption{Abundance patterns of PISN yields normalized by magnesium yield for
$M(\rm He)$ = 113 M$_\odot$ (red points) with the uncertainties from different
$\sigma_{C12\alpha}$ (black triangles) and different $f_{\rm 16O}$ (green triangles). 
The thick end of the triangle shows the results of $\sigma_{C12\alpha} = 1$ or
$f_{\rm 16O} = 10$, while the thin end of the triangle shows the results
of $\sigma_{C12\alpha} = -1$ or $f_{\rm 16O} = 0.1$.
\label{fig:pisn_sigma}}
\end{figure}

In order to reconciling the discrepancies between
the Pop III PISNe models
and the observed abundances from VMP star, it is necessary to quantify the
sensitivity of these abundance patterns to key nuclear reaction rates.
In Figure \ref{fig:pisn_sigma}, we show the abundance patterns predicted for
He113 models (red points) along with the uncertainties arising from the
$^{12}$C($\alpha, \gamma$)$^{16}$O (black triangles) and $^{16}$O+$^{16}$O
(green triangles) reaction rates.
To avoid the complex zigzag patterns in Figure \ref{fig:pisn_yield},
arising from the strong odd-even effect in PISNe, we plot odd-Z and even-Z
elements separately in the subsequent Figures to enhance readability.
 The thick end of the triangle shows 
the result of $\sigma_{C12\alpha} = 1$ or $f_{\rm 16O} = 10$,
while the thin end of the triangle shows the result of
$\sigma_{C12\alpha} = -1$ or $f_{\rm 16O} = 0.1$.

For the intermediate-mass elements ($A$ = 16 -- 40),
as $\sigma_{C12\alpha}$ varies within $\pm 1 \sigma$,
the predicted differences in even-Z elements remain below 0.16 dex.
When $f_{\rm 16O}$ varies over 2 orders of magnitude (from 0.1 to 10), 
the changes in these abundances are smaller than 0.14 dex.
Thus, the even-Z elemental abundances are largely insensitive to
variations in these two reaction rates, because the yields of these 
$\alpha$ elements are mainly proportional to $M(\rm He)$.

Within the uncertainty range of $\sigma_{C12\alpha}$ ($f_{\rm 16O}$),
the predicted variations are [Na/Mg] = 0.6 (0.1), [Al/Mg] = 0.3 (0.05),
[P/Mg] = 0.05 (0.38), [Cl/Mg] = 0.57 (0.38), and [K/Mg] = 0.38 (0.5). 
Notably, the [Cl/Mg] and [K/Mg] abundances predicted for
both $f_{\rm 16O}=$ 0.1 and 10 are higher than
those predicted by $f_{\rm 16O}=$ 0. Thus, 
the uncertainties are evaluated between $f_{\rm 16O}=$ 0 and 10.
The abundances [Na/Mg] and [Al/Mg] are negatively correlated with
$\sigma_{C12\alpha}$, while [S/Mg] and [K/Mg] exhibit positive correlations.
This occurs because lowering $\sigma_{C12\alpha}$ increases
$X$($^{12}$C) in the CO shells,
where the neutron excess increases via $^{12}$C($^{12}$C, n)$^{24}$Mg
and weak interactions involving $^{20, 21}$Ne \cite{1996snih.book.....A, Woosley_2002},
leading to enhanced production of Na and Al.
In contrast, Cl and K are synthesized in the OSi shell,
where the neutron excess is significantly altered by the explosion.
As shown in Figure \ref{fig:exp_mhe} (d), $\eta$ is proportional
to $\sigma_{C12\alpha}$ during the explosion for $M(\rm He)=$ 113 M$_\odot$. 
The Fe-peak elements are significantly sensitive to
$\sigma_{C12\alpha}$ and $f_{\rm 16O}$, because they are
primarily produced in the NSE process, and these two reaction 
rates significantly alter the maximum temperature during the explosion.

\subsection{Comparison with Observations and Other Models}
\label{sec:observe}

When comparing theoretical predictions with observations,
two critical issues regarding PISN abundances demand our attention.
The first concerns the origin of low [$\alpha$/Fe] abundances observed
in VMP stars \cite{2003ApJ...592..906I, 2022ApJ...931..147L, 2023Natur.618..712X}
and high-redshift galaxies \cite{2021ApJ...913...22K, 2022ApJ...925..111I,
2023ApJ...959..100I, 2024ApJ...976..122N, 2025arXiv250311457N}. 
We find that [Mg/Fe]$<$0 for stars with $M(\rm ZAMS) \geq$ 200 M$_\odot$,
which is lower than the minimum [Mg/Fe] predicted by normal CCSN
\cite{2007ApJ...660..516T, 2010ApJ...724..341H, 2012ApJS..199...38L}.
Consequently, PISNe are considered a promising source for explaining the low [$\alpha$/Fe] 
in VMP stars \cite{2023Natur.618..712X}, although certain bright Hypernovae (BrHNe)
may also yield similarly low [$\alpha$/Fe] ratios \cite{2008ApJ...673.1014U}. 

\begin{figure}[H]
\centering
\begin{minipage}[c]{0.48\textwidth}
\includegraphics [width=85mm]{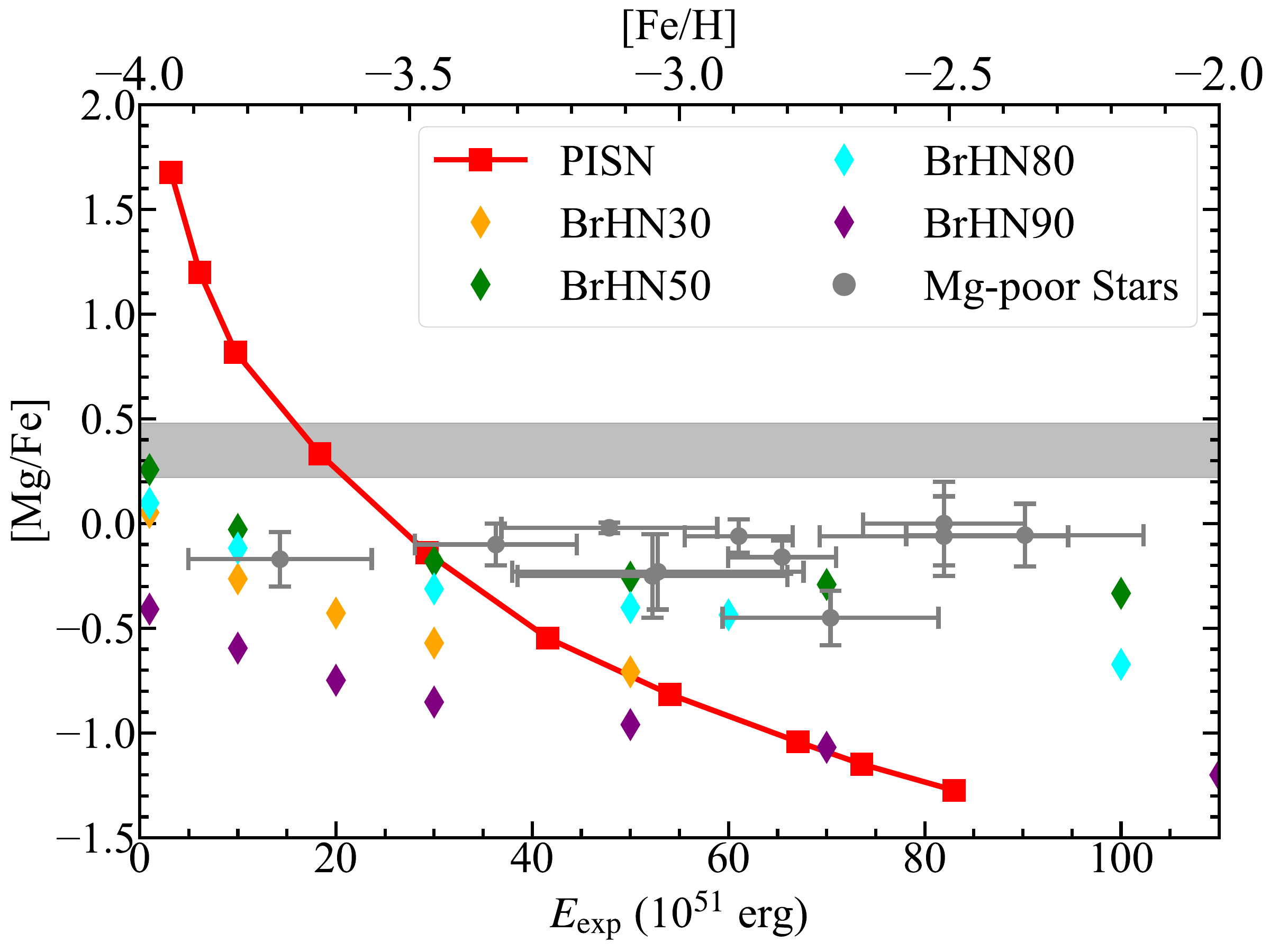}
\end{minipage}%
\caption{Comparison of [Mg/Fe] between VMP stars and theoretical SN models. 
The red squares represent [Mg/Fe] predicted by Pop III PISNe in this work,
while the colored diamonds are [Mg/Fe] predicted by BrHNe from ref. 
\cite{2008ApJ...673.1014U}.
We adopt their upper limit of the ejected $^{56}$Ni, in which the mass cut is
set just above the Fe core. Orange, green, cyan, and purple show the stars
with an initial mass of 30, 50, 80, and 90 M$_\odot$, respectively.
The gray region indicates the observed abundance of 0.42$\pm$0.06 for giant
and 0.29$\pm$0.07 for main-sequence turnoff stars from 400 VMP stars reported in
ref. \cite{2022ApJ...931..147L}. The gray points with error bars represent
the observations of Mg-poor ([Mg/Fe]$<$0) stars in ref. \cite{2022ApJ...931..147L}.
\label{fig:low_alpha}}
\end{figure}

In Figure \ref{fig:low_alpha}, we compare the [Mg/Fe] between the VMP stars
and theoretical SN models.
\cite{2022ApJ...931..147L} reported that for a sample of 400 VMP stars,
[Mg/Fe] clusters around $\sim$0.35 with a small scatter,
represented by the gray region in Figure \ref{fig:low_alpha}.
However, eleven stars exhibit subsolar [Mg/Fe],
and these stars are explicitly marked with error bars.
Both PISNe and BrHNe can produce sufficient Fe to account
for the observed subsolar [Mg/Fe]. 
For the BrHNe models, we adopt the upper limit on the ejected $M(\rm Fe)$ from
\cite{2008ApJ...673.1014U}, in which the mass cut is set just
above the Fe core and no fallback is assumed.
In this study, the [Mg/Fe] ratio predicted by PISNe covers a wide range
from 1.7 to -1.3 for higher $M(\rm He)$.
Only PISNe with $M(\rm ZAMS)=$ 180 - 220 M$_\odot$ can reproduce the observed [Mg/Fe]. 
However, for BrHNe with $M(\rm ZAMS)=$ 30, 50, 80, and 90 M$_\odot$,
explosion energies ($E_{\rm exp}$) of 1--20 foe, 10--100 foe, 1--60 foe,
and 1--10 foe, respectively, are required to match these Mg abundances.

The second challenge lies in distinguishing PISNe from CCSNe to positively identify Pop III imprints.
previous studies have established chemical diagnostics include
low [Na/Mg] ($\sim -1.5$) and high [Ca/Mg] ($\sim 0.5$--$1.3$) \cite{Takahashi_2018_1});
low [N/Fe] ($< 0$) \cite{2019MNRAS.487.4261S};
low [Al/Si] ($< 0.75$) and [Al/Mg] ($< -1.5$) \cite{2024ApJ...967L..22V};
and enhanced [Si,S,Ca/O] ($\sim 0.8$) from explosive O burning \cite{2013ARAA..51..457N}.
While effective for high-mass PISNe,
these diagnostics may overlap with CCSN abundances in the case of low-mass PISNe.
Stars with $M(\rm ZAMS) \leq$ 200 M$_\odot$, corresponding to
$M(\rm He) \leq$ 93 M$_\odot$, explode via incomplete Si burning (Figure \ref{fig:ctrx}).
They produce enhanced [$\alpha$/Fe] and significantly suppressed Ti, Cr, Ni, and Zn.
Therefore, we suggest that the ratios of [(Si, Ca)/(Cr, Ni)] could serve as excellent probes
for distinguishing low mass PISN ($M(\rm He)<$ 93 M$_\odot$) from CCSNe.

\begin{figure}[H]
\centering
\begin{minipage}[c]{0.48\textwidth}
\includegraphics [width=85mm]{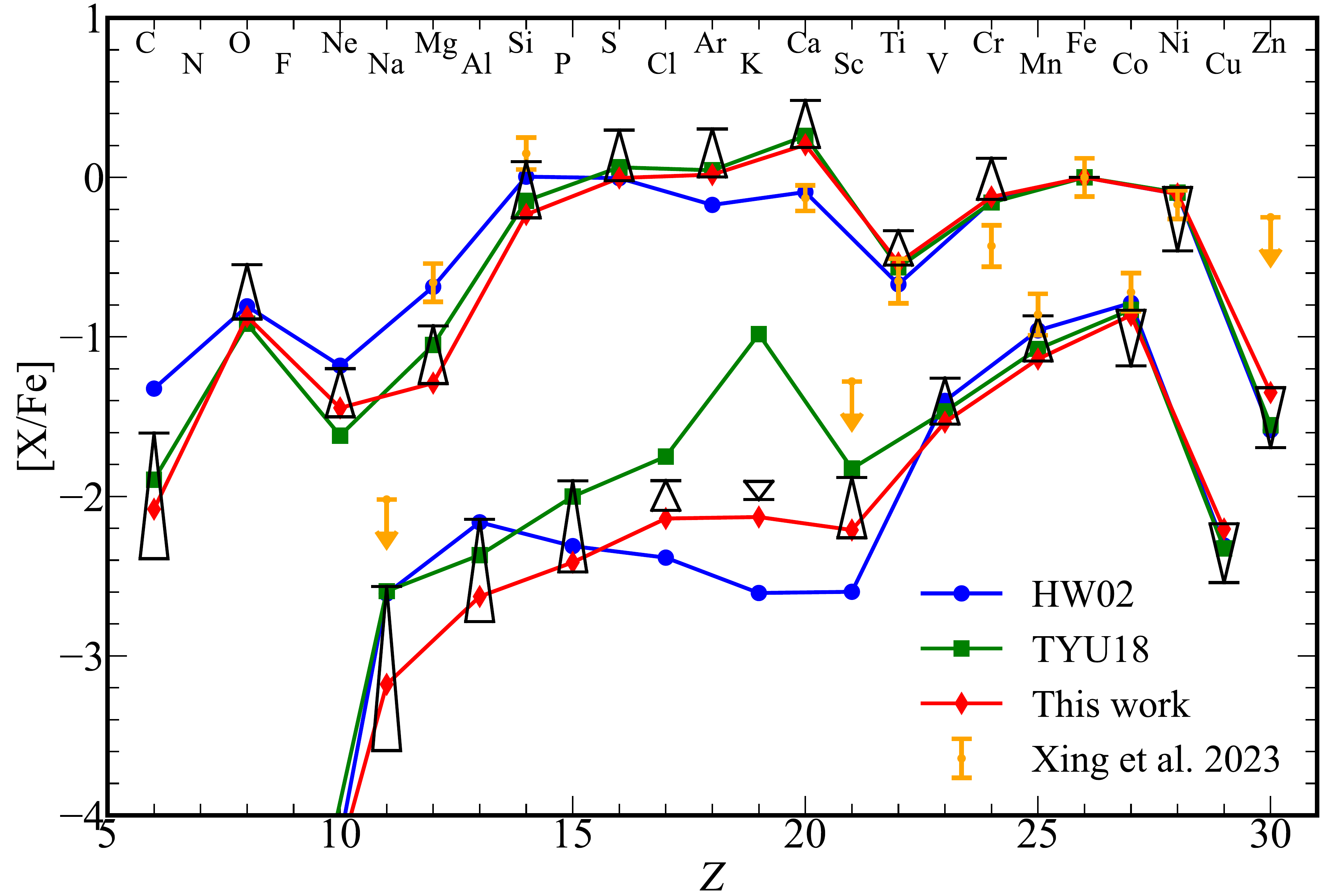}
\end{minipage}%
\caption{Abundances of PISN yields normalized by iron yield for
the most massive PISN in this work, TYU18, and Hw02.
The orange points with the error bars represent the observation of
the VMP star from ref. \cite{2023Natur.618..712X}.
The $M(\rm He)$ of the most massive PISN from these three works
are 133.52, 142.84, and 133 M$_\odot$, respectively.
The triangles indicate the uncertainties from $\sigma_{C12\alpha} = \pm 1$.
Note, for $\sigma_{C12\alpha} = 1$ case, we use the model with $M(\rm He)$
= 128 M$_\odot$ instead of $M(\rm He)$ = 133 M$_\odot$,
because the model with $M(\rm He)$ = 133 M$_\odot$ undergo CC.
\label{fig:pisn_comp}}
\end{figure}

Figure \ref{fig:pisn_comp} compares the elemental abundances predicted by
the most massive PISNe from this study, HW02, and TYU18 with the observation
of LAMOST J1010+2358 \cite{2023Natur.618..712X}. 
$M(\rm He)$ for these three models are 133.52, 130, and 142.84 M$_\odot$,
respectively. while the predicted $M(^{56}$Ni) are 50.4, 39.6, and 47.68 M$_\odot$,
respectively.

For most elements, the predictions from the three models are
consistent well within the uncertainty in $\sigma_{C12\alpha} = \pm 1$.
Given additional uncertainties
in other reaction rates (e.g., $^{16}$O+$^{16}$O, $3 \alpha$),
the total nuclear uncertainties are likely larger.
Although the three models differ in various assumptions,
e.g., the treatment of convection, numerical code, and H-envelope retention,
the predicted final elemental abundances do not vary greatly,
with differences confined primarily to a few specific elements.

For even-$Z$ elements, predictions across models are generally consistent,
with slight deviations observed primarily in [C/Fe],
[Ne/Fe], [Mg/Fe], [Ar/Fe], and [Ca/Fe] from HW02. These discrepancies 
likely arise from the differences in $\alpha$-capture reaction rates.
Conversely, results from this work and TYU18 align closely,
as both studies adopt the same reaction rates from the JINA REACLIB database.
For odd-Z elements, the predictions from this work and HW02 are
relatively similar, as both studies adopted a pure He core and
neglected mixing between the hydrogen envelope and the helium-burning shell.
However, TYU18 retained a hydrogen envelope throughout the evolution.
The resulting mixing of $^{14}$N from the H-rich envelope into
the convective He-burning shell produces additional $^{22}$Ne and releases more neutrons.
Specifically, TYU18 predicts an increasing trend from [Na/Fe] to [K/Fe],
whereas HW02 suggests a gradual decline, with a significant difference of 1.5 dex in [K/Fe].

Under the assumption that LAMOST J1010+2358 is enriched by only a single PISN,
the [Mg/Fe] ratios predicted by this study and TYU18 are underestimated,
while the [Ca/Fe] ratios are overestimated.
The abundances predicted by HW02 align well with LAMOST J1010+2358.
Furthermore, given the current absence of high-precision observational data
for Na, Al, Sc, and Zn, alternative interpretations for LAMOST J1010+2358 remain viable. 
Recent simulations indicate that the abundance pattern of LAMOST J1010+2358 could
potentially be reproduced by low-mass CCSNe \cite{2024MNRAS.527.4790J, 2024OJAp....7E..66T, 
2024OJAp....7E..83J}. Moreover, ref. \cite{2024ApJ...968L..23S} suggested that
LAMOST J1010+2358 could result from a mixed enrichment of Pop III PISNe and Pop II CCSNe. 
Consequently, the chemical signature of J1010+2358 does not strictly
require a single-event enrichment scenario,
and its interpretation as a pure PISN descendant should be treated with caution. 

Owing to the strong odd-even effects and relatively low explosion temperatures,
odd-$Z$ elements (e.g., Na, Al, Sc) and certain Fe-peak elements (e.g., Cu, Zn) serve as
effective diagnostics for discriminating between the PISN and non-PISN enrichment.
However, whether Pop II PISNe share similar nucleosynthetic properties
with their Pop III counterparts remains uncertain.
Furthermore, no previous studies have investigated whether Pop II PISNe could reproduce
the observed abundance pattern of LAMOST J1010+2358. To address this gap, 
we explore nucleosynthesis of Pop II PISNe in Section \ref{sec:popII}. 

\section{Fe-enriched PISNe and the Early Chemical Enrichment} \label{sec:popII}

The initial mass function (IMF) of the first stars is under debate,
making it challenging to define the initial composition of Pop II stars.
Some recent simulations \cite{Abel_1998, 2000ApJ...540...39A, Hirano_2015} have
suggested that the masses of the first stars are heavier than 100 M$_\odot$.
Among these stellar masses, PISNe (140 M$_\odot$ $\leq M(\rm ZAMS) \leq 300$ M$_\odot$)
represent the dominant sources of metal enrichment.
Those with $M(\rm ZAMS) \geq 300$ M$_\odot$ 
are predicted to directly collapse into BHs \cite{2002ApJ...567..532H}.
Furthermore, observations from refs.
\cite{2022ApJ...925..111I, 2022ApJ...937...61Y, 2024ApJ...976..122N}
indicate high Fe abundances in extremely metal-poor galaxies
and high redshift objects, which provide clues to a significant contribution
from PISN ejecta in the early Universe.

Motivated by these studies, we propose an extreme astrophysical scenario 
to trace the contribution of Pop III PISNe in the early Universe.
We assume that in an early galaxy characterized by an extremely top-heavy IMF,
one or several very massive Pop III stars end their brief lives as PISNe.
Within a relatively short timescale following these explosions,
and within a highly localized spatial surroundings,
the primordial gas is enriched exclusively by the PISNe ejecta.
Crucially, the next generation of massive stars born
before any other types of supernova products—such as those from CCSNe,
faint SNe, or Hypernovae—have sufficient time to diffuse into this region.
In Section \ref{sec:imf}, we describe the initial composition models and
assumptions for Pop II PISNe. In Section \ref{sec:evo_exp_popII} and
\ref{sec:nuc_PopII}, the evolution and Nucleosynthesis of Fe-enriched
PINSe are compared with Pop III PISNe, respectively.
We discuss the abundance pattern predicted by different studies and different
initial compositions in Section \ref{sec:diss}.
Finally, the implications for chemical enrichment are discussed in Section \ref{sec:CE}.

\subsection{The Chemical Distribution enriched by Pop III PISNe} \label{sec:imf}

We assume a first galaxy with the total gas mass of  $M_{\rm g}$
and most of the stars formed in this galaxy exceed 140 M$_\odot$.
Since the stars with $M(\rm ZAMS)$ larger than 300 M$_\odot$ collapse into BHs
directly, this galaxy would be enriched by only PISNe.
We also assume that the total mass of exploded stars is $M_*$.
Then, the parameter $f_* = M_*/M_{\rm g}$ is the product of two free parameters:
$f_{\rm form}$ and $f_{\rm exp}$. $f_{\rm form}$ represents the ability for
converting the gas into stars in this first galaxy, which depends on the
intrinsic properties of the galaxy and is largely uncertain.
$f_{\rm exp}$ represents the ratio between the total mass of the exploded stars
and the total stellar mass.

\begin{figure}[H]
\centering
\begin{minipage}[c]{0.48\textwidth}
\includegraphics [width=85mm]{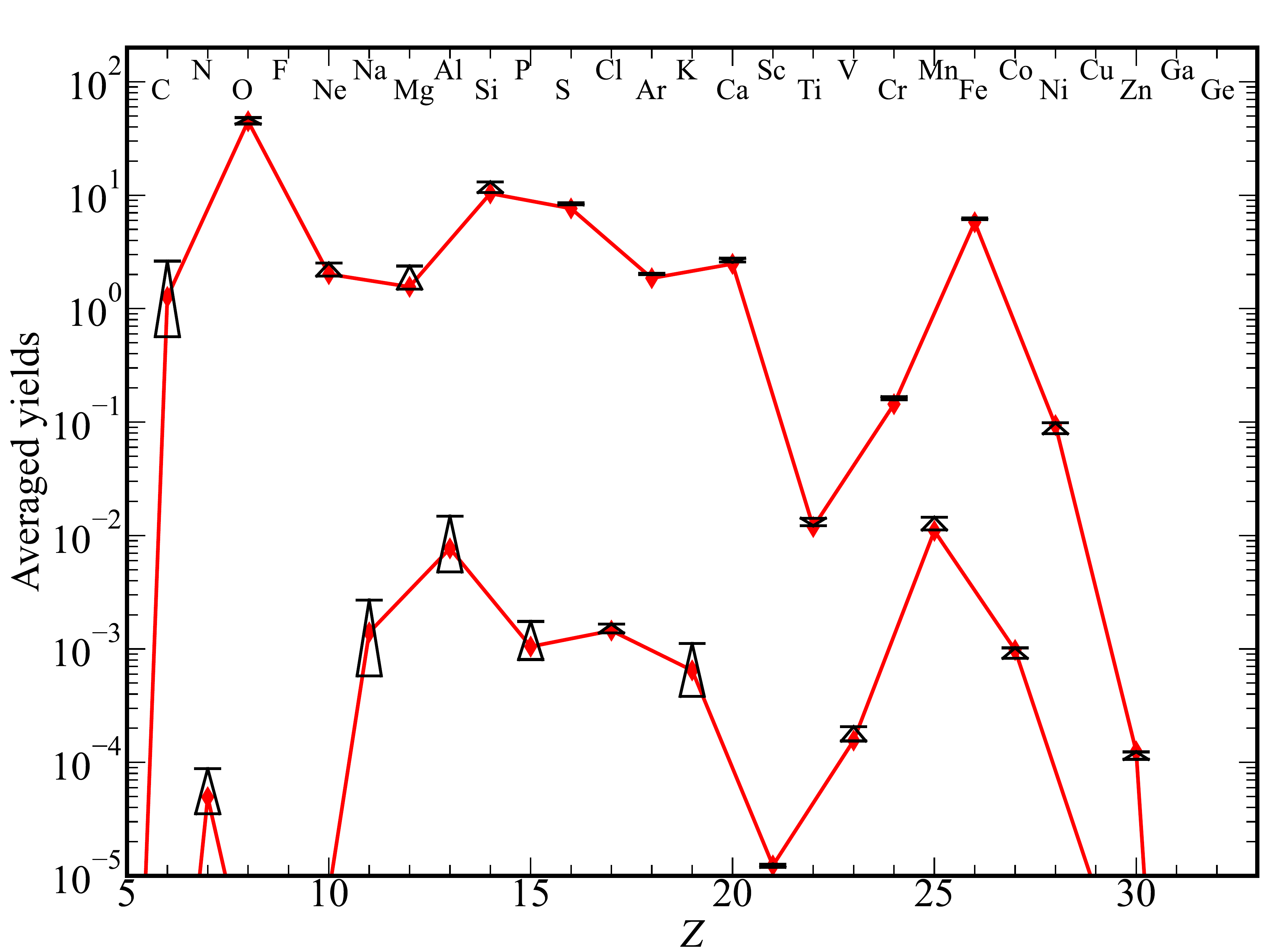}
\end{minipage}%
\caption{The predicted yields of Pop III PISNe from $M(\rm ZAMS)=$ 140 - 280 M$_\odot$
integrated over a Salpeter-like IMF with the slope of $\gamma=-2.35$.
The red points represent the $\sigma_{C12\alpha}=$ 0 case.
The uncertainties from different $\sigma_{C12\alpha}$ is shown by the triangles.
The thick end of the triangle shows the results of $\sigma_{C12\alpha} = 1$,
while the thin end of the triangle shows the results of $\sigma_{C12\alpha} = -1$.
\label{fig:pisn_imf}}
\end{figure}

Then, we can calculate the total mass of each isotope $i$ ejected after
these first stars explode:
\begin{equation} \label{equ:mass_i}
M_i = X_i M_*
\end{equation}
where $X_i$ represents the average yield of isotope $i$ synthesized by per stellar mass
and is defined as:
\begin{equation} \label{equ:y_star}
X_i = \frac{Y_i}{\sum_k Y_k}
\end{equation}
where $Y_i$ represents the yield of isotope $i$ integrated by a Salpeter-like IMF
across the initial masses of PISNe and $\sum_k Y_k$ runs over all isotopes.
Since the initial mass range we studied is limited, the integration results
are not sensitive to the slope of IMF \cite{2002ApJ...567..532H}.
Figure \ref{fig:pisn_imf} shows that most $Y_i$s are also not sensitive
to the $\sigma_{C12\alpha}$, except C, N, Na, Al, P, and K. 
Therefore, only $\sigma_{C12\alpha}=$ 0 is considered hereafter.

Similarly, we can estimate roughly the metallicity of the ISM after several
PISNe in this galaxy by:
\begin{equation} \label{equ:mz1}
Z_{\rm ISM} = \frac{\sum_j M_i}{M_{\rm g}}
\end{equation}
Note $\sum_j M_i$ runs over all the isotopes heavier than $^4$He.
$M_{\rm g}$ is the total gas mass in the galaxy.
However, not all these fresh metals will necessarily to retained by the galaxy
and not all the gas will be used to dilute metals.
Because the mechanism of the metal mixing process is still uncertain,
we use a free parameter $f_{\rm dil}$ and Equation \ref{equ:mz1} can be written as
\begin{equation} \label{equ:mz2}
Z_{\rm ISM} = \frac{\sum_j M_i}{f_{\rm dil} M_{\rm g}}
\end{equation}
Combining with Equation \ref{equ:mass_i} and $f_* = M_*/M_{\rm g}$,
we can obtain:
\begin{equation} \label{equ:mz3}
Z_{\rm ISM} = \frac{f_* }{f_{\rm dil}} \sum_j X_i
\end{equation}
Also, we can calculate the abundance of [Fe/H] in the ISM as:
\begin{equation} \label{equ:fe_h}
\begin{split}
[{\rm Fe/H}]_{\rm ISM} &\equiv {\rm log }[\frac{N_{\rm Fe}}{N_{H}}] - {\rm log }[\frac{N_{\rm Fe}}{N_{H}}]_\odot \\
                       &\simeq {\rm log }[\frac{M_{\rm Fe}}{M_{H}}] - {\rm log }[\frac{M_{\rm Fe}}{M_{H}}]_\odot \\
                       &= {\rm log }[\frac{\sum_{\rm Fe} M_i}{M_{\rm dil}}] - {\rm log }[\frac{M_{\rm Fe}}{M_{H}}]_\odot \\
\end{split}
\end{equation}
where $\sum_{\rm Fe} M_i$ runs over all the isotopes of Fe.
$M_{\rm dil} = f_{\rm dil}/f_* M_{\rm H}$ represents the mass of H,
which is provided by primordial gas and would be mixed effectively
with ejected metals, if we assume all the ejecta are retained in the galaxy.
More details about the free parameters $f_{\rm dil}/f_*$ have been
discussed by ref. \cite{2019MNRAS.487.4261S}.

\begin{table}[H]
\centering
\caption{The primordial gas mass effectively mixed with the ejecta ($M_{\rm dil}$),
metallicity ($Z$) for various [Fe/H].
The composition of primordial gas is assumed to have originated from the Big Bang
nucleosynthesis from ref. \cite{2007ARNPS..57..463S}.}
\label{tab:my-metal}
\begin{tabular}{ccc}
\hline
\hline
[Fe/H]  & $M_{\rm dil}$ (M$_\odot$) &   $Z$                  \\
\hline
0       & 4.13$\times 10^3$         & 0.0186                 \\
$-1$    & 4.13$\times 10^4$         & 1.86$\times 10^{-3}$   \\
$-2$    & 4.13$\times 10^5$         & 1.86$\times 10^{-4}$   \\
$-4$    & 4.13$\times 10^7$         & 1.86$\times 10^{-6}$   \\
$-6$    & 4.13$\times 10^9$         & 2.09$\times 10^{-8}$  \\
\hline
\hline
\end{tabular}
\end{table}

\begin{figure}[H]
\centering
\begin{minipage}[c]{0.48\textwidth}
\includegraphics [width=85mm]{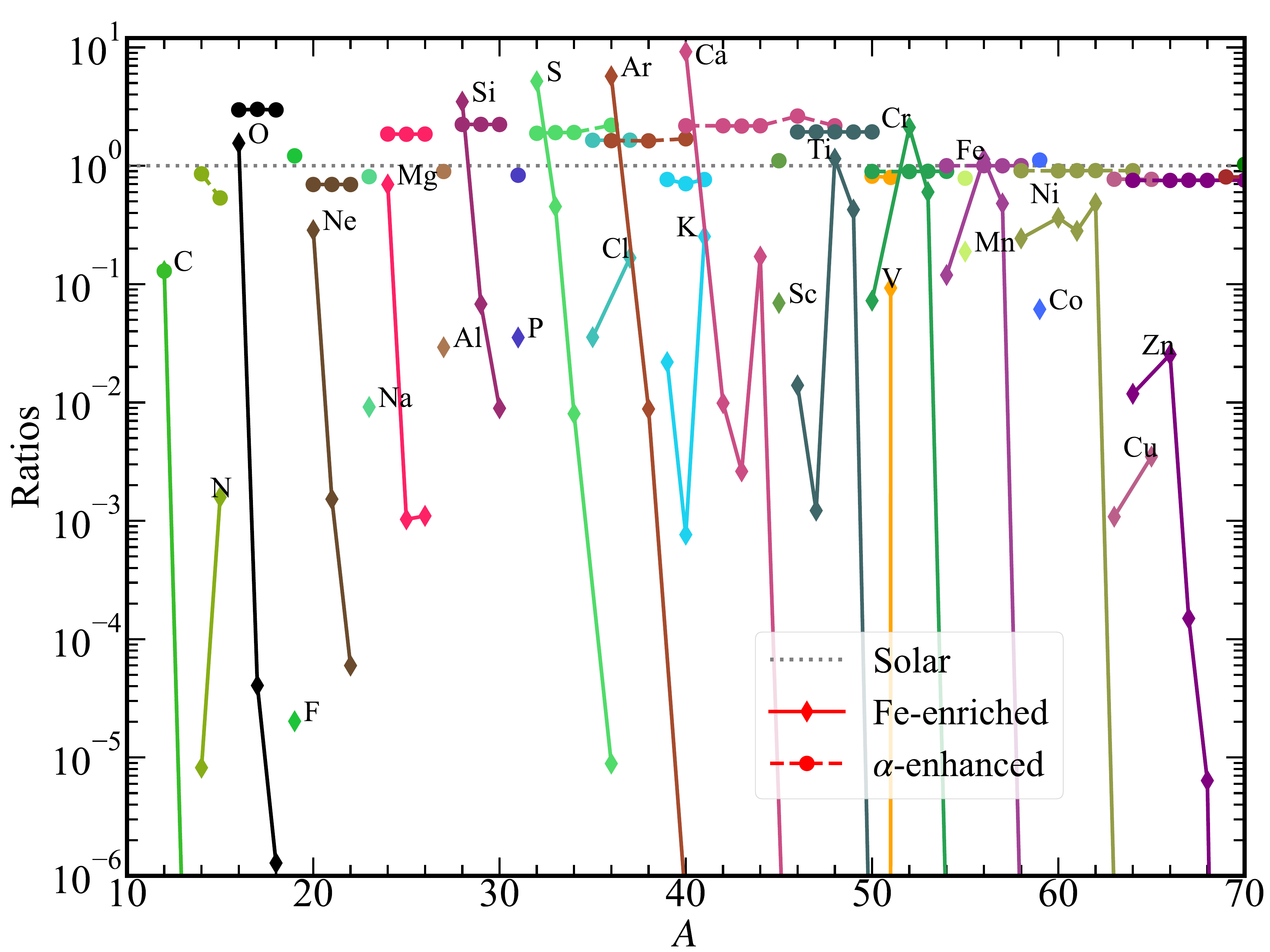}
\end{minipage}%
\caption{Comparison of the different initial compositions for metal-poor
stars with [Fe/H] = -1. The Fe-enriched composition is adopted in this study.
The exceptions of [$\alpha$/Fe] ratios are enhanced to fit the observations
from metal-poor stars in the $\alpha$-enhanced composition.
Both compositions are normalized by 0.1 Z$_\odot$.
\label{fig:solar_ratio}}
\end{figure}

Here, $\sum_k Y_i=$ 84.4 M$_\odot$ gives the total mass of ejecta.
After PISNe explosions, we assume that [Fe/H]$_{\rm ISM}$ reaches
$-$6, $-$4, $-$2, $-$1 and even 0. Then, according to Equation \ref{equ:fe_h},
we can easily calculate the primordial gas mass effectively mixed with the ejecta
($M_{\rm dil}$) and the mass fraction of each isotope.
We show $M_{\rm dil}$ and metallicity ($Z$) in Table \ref{tab:my-metal}.

In Figure \ref{fig:solar_ratio}, we compare the subsolar isotopic
compositions adopted in different studies.
For [Fe/H] = -1, the total metallicities in Fe-enriched, solar,
and $\alpha$-enhanced compositions are 1.86$\times$10$^{-3}$,
1.41$\times$10$^{-3}$ and 3.236$\times$10$^{-3}$, respectively.

Previous Pop II PISNe models \cite{2014AA...566A.146K, 2024ApJ...961..146U}
adopted a scaled fraction of solar composition as initial composition
for the [Fe/H] = -1. However, the solar mixture is not
entirely realistic, as it includes contributions from Type Ia supernovae.
The $\alpha$-enhanced composition from \cite{2018ApJS..237...13L}
provides a more realistic alternative, as the [$\alpha$/Fe] ratios are
specifically enhanced to match observations of metal-poor stars.
The Fe-enriched composition adopted in this study represents an extreme case.

Compared with the solar abundance, the $^{16}$O, $^{28}$Si, $^{32}$S, $^{36}$Ar,
$^{40}$Ca, $^{48}$Ti, and $^{52}$Cr are enhanced in Fe-enriched composition.
The neutron-rich isotopes are significantly reduced because of
the high $Y_{\rm e}$ in Pop III PISNe (see Figure \ref{fig:exp_mhe} (d)).
Particularly, the mass fraction of $^{14}$N is 5 orders of magnitude lower than
that in the solar and $\alpha$-enriched compositions,
This implies that our Pop II models experience limited enhancement of the
neutron excess during the He-burning phase.
A detailed discussion on the effects of adopting different initial compositions
is provided in Section \ref{sec:diss}.

\subsection{The Evolution and Explosion of Pop II PISNe} \label{sec:evo_exp_popII}

At [Fe/H] = 0, stellar wind mass loss heavily constrains the He core mass,
rendering it too small to explode as PISNe.
At [Fe/H] = -1, although mass loss is somewhat weaker,
it can still modestly reduce the He core mass. The results depend on
the adopted wind scheme and whether rotational effects are included,
as discussed in refs. \cite{2007AA...475L..19L, 2011MNRAS.412L..78Y,
2024ApJ...961..146U}.
In this study, we follow the \texttt{Dutch} mass loss prescriptions,
in which the mass loss rate is employed
from ref. \cite{1988AAS...72..259D} for cool stars,
ref. \cite{2001AA...369..574V} for hot hydrogen-rich stars,
and ref. \cite{2000AA...360..227N} for Wolf-Rayet stars, 
and the scaling factor from ref. \cite{2001AA...373..555M}.

\begin{table}[H]
\centering
\caption{Summary table for the H-rich Models with different initial mass and [Fe/H].
$M(\rm final)$ and $M(\rm He)$ represent the star mass and the He core mass at the end
of He burning. $Z_{\rm He}$ represents the mass fraction of metal at the He ignition,
which is used as the initial metallicity of the He star model.}
\begin{tabular}{ccccc}
\hline
\hline
[Fe/H] & $M(\rm ZAMS)$ & $M(\rm final)$  & $M(\rm He)$  & $Z_{\rm He}$ \\
       &  (M$_\odot$)  & (M$_\odot$)     & (M$_\odot$)  &    \\
\hline
$-\infty$ & 270 & 270.0 & 128.19 & 4.4$\times 10^{-6}$ \\ 
$-$6 & 270 & 270.0 & 127.58 & 4.7$\times 10^{-6}$ \\ 
$-$4 & 257 & 256.9 & 127.75 & 4.0$\times 10^{-4}$ \\ 
$-$2 & 243 & 239.6 & 128.39 & 1.0$\times 10^{-3}$ \\ 
$-$1 & 244 & 211.1 & 127.54 & 3.2$\times 10^{-3}$ \\
\hline
\hline
\end{tabular}
\label{tab:my-popii}
\end{table}

In addition to mass loss, overshooting is strengthened
with increasing metallicity \cite{2013ApJ...767...78T},
which results in a larger $M(\rm He)$.
Since PISNe explosion properties are highly sensitive to $M(\rm He)$
(see Figure \ref{fig:exp_mhe}),
we vary the initial masses to maintain roughly consistent
$M(\rm He)$ across all metallicities.
In this section, we use an example of $M(\rm He)$ = 128 M$_\odot$
to investigate the effect of metallicity on the PISN explosion and
associated nucleosynthesis.
Table \ref{tab:my-popii} presents the initial mass $M(\rm ZAMS)$,
He core mass $M(\rm He)$, and the star mass at the end of 
the He burning $M(\rm final)$.
$Z_{\rm He}$ represents the mass fraction of metal at the He ignition,
which is used as the initial metallicity of the He star model.
For [Fe/H] $\leq -6$, $Z_{\rm He}$ does not exhibit any significant
change, and the mass loss can essentially be neglected.
Consequently, the subsequent evolution of the progenitors and
their explosions should be similar for [Fe/H] $\leq -6$.
In the subsequent sections,
we focus on $M(\rm He)$ = 83 and 128 M$_\odot$ (hereafter He83 and He128)
with [Fe/H] = $-$1 and $-$2. The properties for these models are summarized
in Table \ref{tab:my-pisn}.

\begin{figure*}[]
\centering
\begin{minipage}[c]{0.9\textwidth}
\includegraphics [width=160mm]{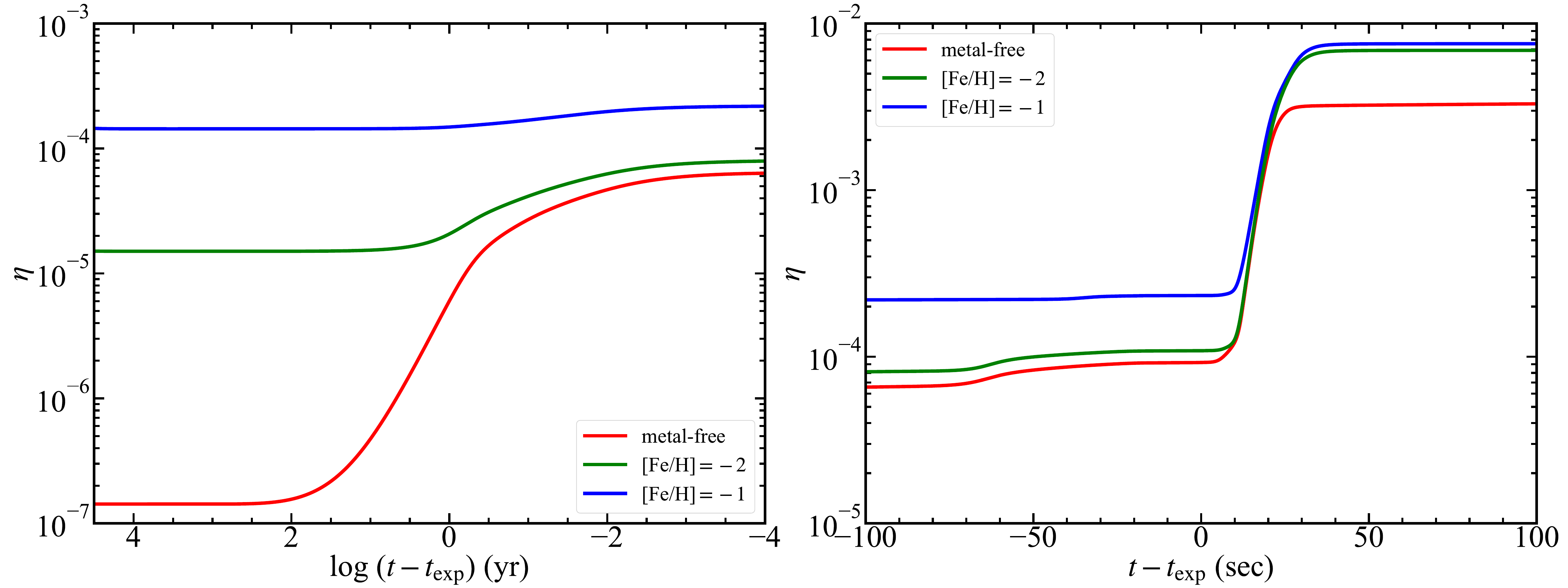}
\end{minipage}%
\caption{The time evolution of neutron excess ($\eta$) at the center of stars
from the ignition of He burning to the onset of explosive O burning for
$M(\rm He)$ = 128 M$_\odot$ but various metallicities.
$t_{\rm exp}$ represents the time when $T_{\rm c} = 10^{9.5}$ K.
The top panel shows the evolution from core He burning to $10^3$ s
before the explosion. The bottom panel shows the evolution within 200 s
during the explosion.
\label{fig:eta_metal}}
\end{figure*}

\begin{figure*}[htbp]
\centering
\begin{minipage}[c]{0.9\textwidth}
\includegraphics [width=160mm]{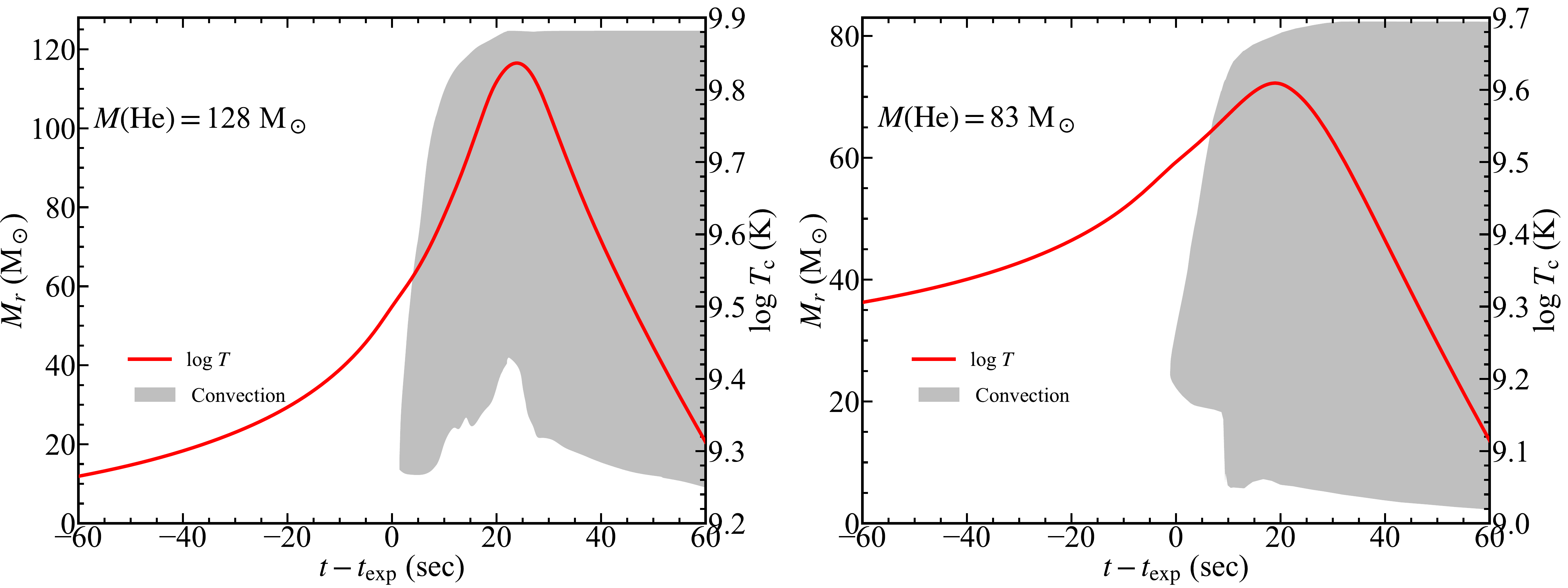}
\end{minipage}%
\caption{The time evolution of convection regions and central temperature
near the explosion for [Fe/H] = -1 but $M(\rm He)$ = 83 (left) and 128 (right)
M$_\odot$, respectively.
$t_{\rm exp}$ represents the time when $T_{\rm c} = 10^{9.5}$ K.
\label{fig:kipp}}
\end{figure*}

Figure \ref{fig:eta_metal} displays the time evolution of neutron excess
($\eta$) for the He128 model and $t_{\rm exp}$ is 
defined at $T_{\rm c}$ = 10$^{9.5}$~K,
corresponding to the onset of explosive O burning.
In the top panel, no enhancement in neutron excess is observed
during core He burning ($t-t_{\rm exp}>10^{3.3}$ yr).
This is primarily because of the deficiency of the initial $^{14}$N,
as previously mentioned.
After core He burning, C burns radiatively both centrally and
off-center, due to the limited availability of fuel ($X$($^{12}$C)$<$ 0.05). 
In the absence of convection, fresh fuel cannot be transported
from the outer layers to the inner regions.
Consequently, as shown in Figure \ref{fig:kipp},
the CO core contracts and temperature increases quickly
\cite{Farmer_2020, 2023RAA....23a5014X}.
Reactions such as $^{12}$C($^{12}$C, n)$^{23}$Mg($\beta^-$)$^{23}$Na
and $^{20}$Ne(p,$\gamma$)$^{21}$Na 
($\beta^-$)$^{21}$Ne(p,$\gamma$)$^{22}$Na($\beta^-$)$^{22}$Ne
drive a limited increase in neutron excess before the explosion.
As the [Fe/H] increases, the enhancements gradually weaken.
In the bottom panel, when the explosive O burning is initiated,
the neutron excess remains below $10^{-3}$,
which is the typical value of solar metallicity massive
stars at core helium exhaustion.
We find that a significant enhancement in neutron excess occurs after
$t_{\rm exp}$ = 10 s.
The neutron excesses in all metallicities increase by more than 1
order of magnitude. For stars with [Fe/H]$=-1$, the neutron excess
reaches 7.6$\times10^{-3}$, slightly higher than that of metal-free models.
As a result, $X$($^{56}$)Ni is reduced slightly
while neutron-rich isotopes are formed in the center for [Fe/H]$=-1$,
particularly $X$($^{54}$Fe) = 6.21 $\times 10^{-2}$.

Such differences originate from the distinct behaviors of outer shells.
In the case of metal-free models, the opacity is very low,
allowing energy to be transported to the surface with little hindrance.
In contracts, for [Fe/H]$=-1$ models, the higher opacity results in
more energy absorbed by the stellar material.
Consequently, a substantial convection zone, indicated by the gray-shaded region 
in Figure \ref{fig:kipp} for both He83 (left) and He128 (right) models,
is triggered at the mass coordinate of $M_r \simeq$ 23 and 13 M$_\odot$
for He83 and He128 models, respectively, when $t \simeq t_{\exp}$.
Then, the top of the convective region rapidly extends outward to the outer
layers within 25 seconds.
After the explosion, the base of the convective region also moves inward to
$M_r \simeq$ 4 M$_\odot$.

\begin{table*}[htbp]
\centering
\caption{Summary table for the PISN models with $M(\rm He)$ = 83 and 128 M$_\odot$
but various metallicities. 
$M(\rm He)$ and $X$($^{12}$C) are the He core mass and central mass fraction of $^{12}$C
at the end of core He burning.
$M$($^{56}$Ni) and $E_{\rm tot}$ are the ejected $^{56}$Ni mass and total energy.
log $\rho_{\rm max}$, log $T_{\rm max}$ and $\eta$ are maximum central density
and temperature and the neutron excess during the explosion.}
\begin{tabular}{ccccccccc}
\hline
\hline
[Fe/H] & $M(\rm He)$ &  $X$($^{12}$C) & $M(^{56}$Ni) & $E_{\rm tot}$   & log $\rho_{\rm max}$ & log $T_{\rm max}$ &  $\eta$\\
       & (M$_\odot$) &         & (M$_\odot$)  & ($10^{52}$ erg) & (g cm$^{-3}$)        & (K)               &             \\
\hline
$-\infty$ & 128 &  0.0492 & 38.7   & 7.36  &  6.91  &  9.78 &  3.44$\times 10^{-3}$  \\ 
$-$2      & 128 &  0.0484 & 63.9   & 8.92  &  7.06  &  9.82 &  6.91$\times 10^{-3}$  \\ 
$-$1      & 128 &  0.0486 & 67.3   & 9.02  &  7.15  &  9.84 &  7.56$\times 10^{-3}$  \\
\hline
$-\infty$ & 83  &  0.0643 & 1.39   & 1.84  &  6.38  &  9.60 &  3.04$\times 10^{-4}$  \\ 
$-$2      & 83  &  0.0617 & 0.96   & 2.60  &  6.41  &  9.61 &  3.80$\times 10^{-4}$  \\ 
$-$1      & 83  &  0.0623 & 0.79   & 2.49  &  6.41  &  9.61 &  6.84$\times 10^{-4}$  \\
\hline
\hline
\end{tabular}
\label{tab:my-pisn}
\end{table*}

\begin{figure}[H]
\centering
\begin{minipage}[c]{0.48\textwidth}
\includegraphics [width=85mm]{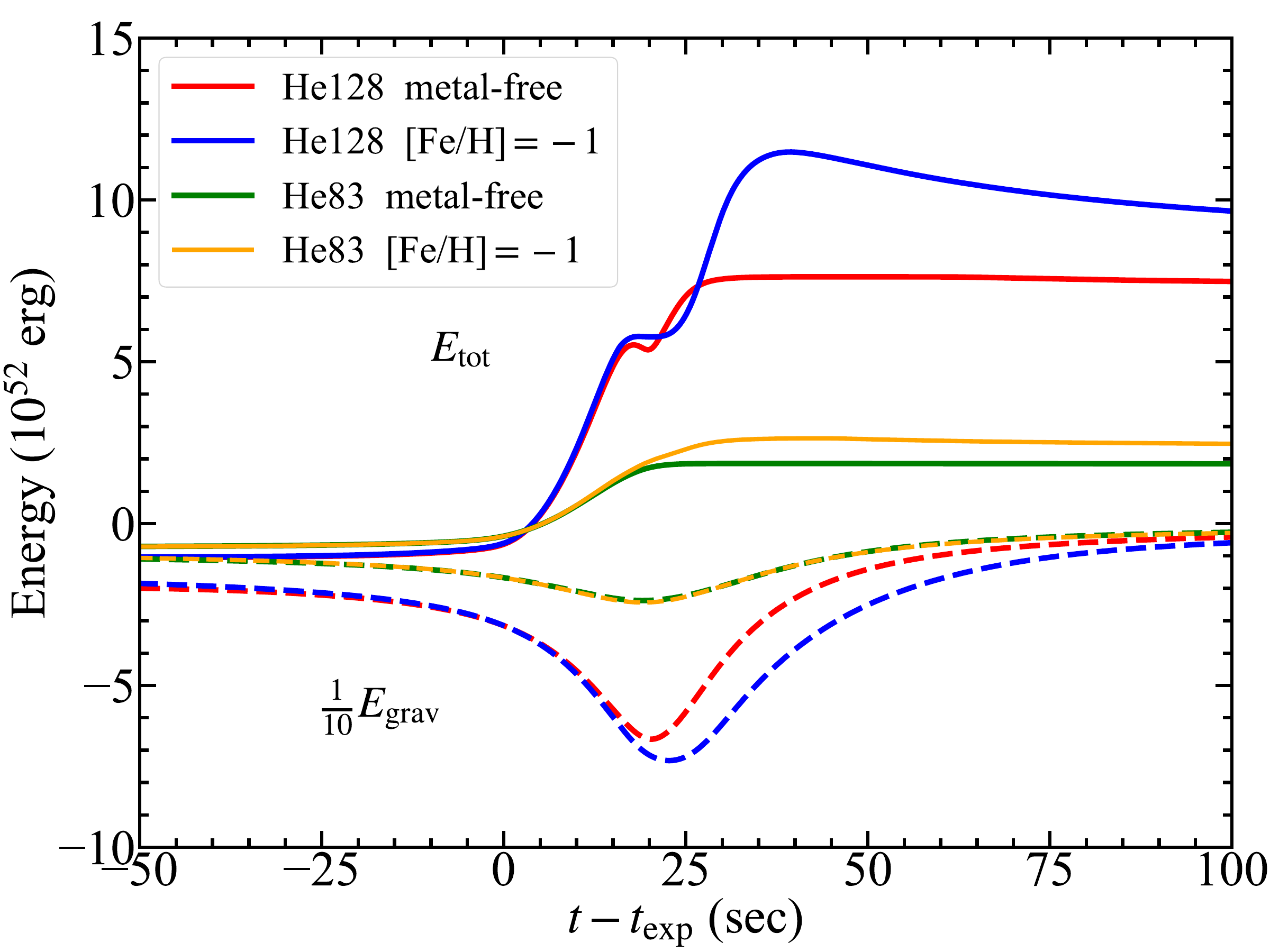}
\end{minipage}%
\caption{The time evolution of the energetics for metal-free and [Fe/H]$=-1$ models.
The total energy and gravitational energy are represented by solid and dashed lines,
respectively. He83 and He128 indicate the models with He core masses of $M(\rm He)$
= 83 and 128 M$_\odot$.
$t_{\rm exp}$ represents the time when $T_{\rm c} = 10^{9.5}$ K.
\label{fig:energy_time}}
\end{figure} 

Convection occurring in stars with [Fe/H] = $-$1 enhances the
efficiency of energy transport generated in the core outward.
This results in more gravitational energy being released
through the contraction to compensate for energy losses,
as illustrated in Figure \ref{fig:energy_time}.
Table \ref{tab:my-pisn} indicates that these stars reach higher central
temperatures and densities compared to metal-free stars.
Additionally, we find that the total energies, $E_{\rm tot}$, 
are identical for both metal-free and [Fe/H] = $-$1 cases
until the stars reach the highest central temperature.
Beyond this point, $E_{\rm tot}$ continues to rise in [Fe/H] = $-$1 stars,
while it stays approximately constant in the metal-free counterparts.
We also find that Pop II PISNe are energetic events,
with explosion energies reaching 10$^{53}$ erg (approximately
100 times that of normal CCSNe) and $^{56}$Ni up to 76 M$_\odot$.
making them viable candidates for explaining superluminous SNe.

\subsection{The Nucleosynthesis of Pop II PISNe} \label{sec:nuc_PopII}

\begin{figure*}[htbp]
\centering
\begin{minipage}[c]{0.9\textwidth}
\includegraphics [width=160mm]{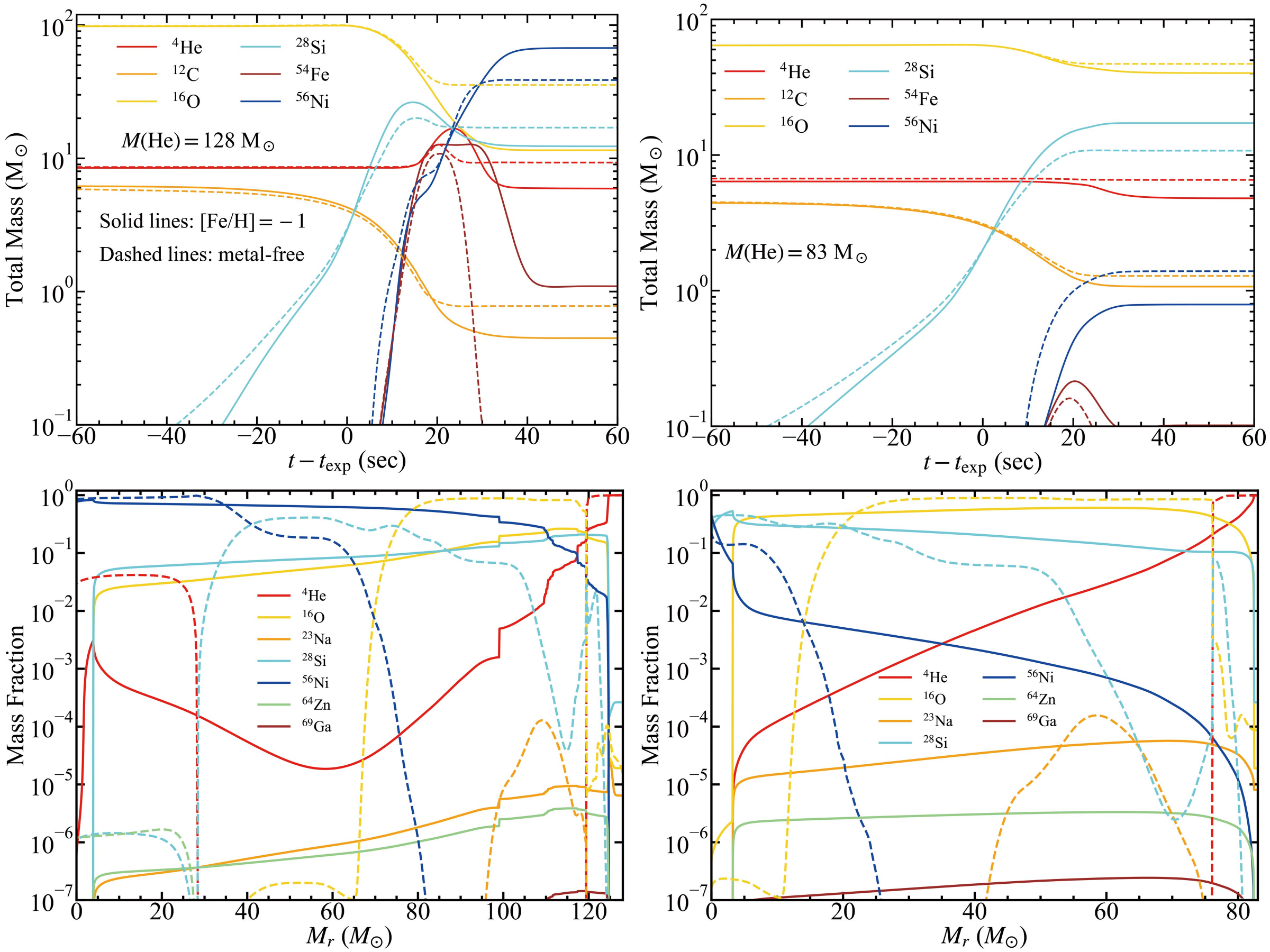}
\end{minipage}%
\caption{Top: The time evolution of the total mass of
some isotopes in the Pop III stars (dashed lines)
and the [Fe/H]$=-1$ stars (solid lines).
$t_{\rm exp}$ represents the time when $T_{\rm c} = 10^{9.5}$ K.
Bottom: The chemical distributions after the PISNe explosion
in the Pop III stars (dashed lines) and the [Fe/H]$=-1$ stars (solid lines).
Note that the left column shows $M(\rm He)$ = 128 M$_\odot$ and
the right column shows $M(\rm He)$ = 83 M$_\odot$.
\label{fig:isos_evo}}
\end{figure*}    

As depicted in the top row of Figure \ref{fig:isos_evo},
the total mass of $^{4}$He in the He128 model decreases from 16.76 M$_\odot$
to 5.91 M$_\odot$ in stars with [Fe/H] = $-$1, while a reduction from 12.7 to
9.3 M$_\odot$ in their metal-free counterparts is observed.
Consequently, an additional 7.45 M$_\odot$ of $^{4}$He is consumed in the
[Fe/H] = -1 model relative to the metal-free one.
This excess $^{4}$He recombines into $^{56}$Ni,
and the energy released during this process primarily drives the continued
increase in $E_{\rm tot}$.
Two reasons account for the enhanced consumption of $^{4}$He.
First, the higher central temperatures in the [Fe/H] = $-$1 stars
tear more $^{56}$Ni into $^{4}$He, thereby depositing additional energy.
Second, convective mixing during the recombination phase transports
$^{4}$He from the outer layers into the inner regions,
facilitating further consumption and energy release. Concurrently, 
a greater mass of $^{56}$Ni are synthesized in the [Fe/H] = $-$1 models.
For the He83 model, the situation is somewhat different.
Since both the metal-free and [Fe/H] = $-$1 progenitors explode before the onset of Si burning.
Although the [Fe/H] = $-$1 stars consume more $^{4}$He, $^{12}$C, and $^{16}$O, thus resulting
in a higher $E_{\rm tot}$, the final $^{56}$Ni mass is lower
because $^{28}$Si remains the dominant product during this phase.

The bottom row of Figure \ref{fig:isos_evo} compares the chemical distributions
between two metallicities at the moment all ejecta become unbound.
In the metal-free He128 model, NSE products such as $^{64}$Zn,
along with residual $^{4}$He, are concentrated in the core ($M_r \leq 30$ M$_\odot$).
However, convection distributes both $^{4}$He and $^{64}$Zn throughout the star
in the [Fe/H] = -1 models. We also note that $^{64}$Zn within $M_r \leq5$ M$_\odot$
is destroyed by the higher temperature in the center.
An enhanced production of nuclei in the Zn--Ge region is seen in the [Fe/H] = $-$1 models.
Since the nuclear network in this work is truncated at Ge, it remains difficult to definitively
determine whether this enhancement originates from the NSE process or the s-process.
Furthermore, light isotopes including $^{16}$O and $^{23}$Na reside within the CO shells
in the metal-free models but are transported inward in the [Fe/H] = $-$1 cases.
The He83 models explode before the onset of Si burning,
reaching peak temperatures far below those required for the NSE process.
Nevertheless, significant yields of $^{64}$Zn and $^{64}$Ge still exist
throughout the ejecta.
These elements may potentially be synthesized via alternative pathways,
possibly involving the s-process.
In future studies, a nuclear network extending beyond the Ge boundary
to the light s-process peak ($Z=$ 38--40; Sr--Y--Zr) is required.
This extension is essential to conclusively identify the nucleosynthetic pathways
responsible for the Zn--Ge enhancement
and to fully decouple the chemical signatures of the NSE process from 
actual weak s-process paths under neutron-deficient conditions.

\begin{figure*}[htbp]
\centering
\begin{minipage}[c]{0.9\textwidth}
\includegraphics [width=160mm]{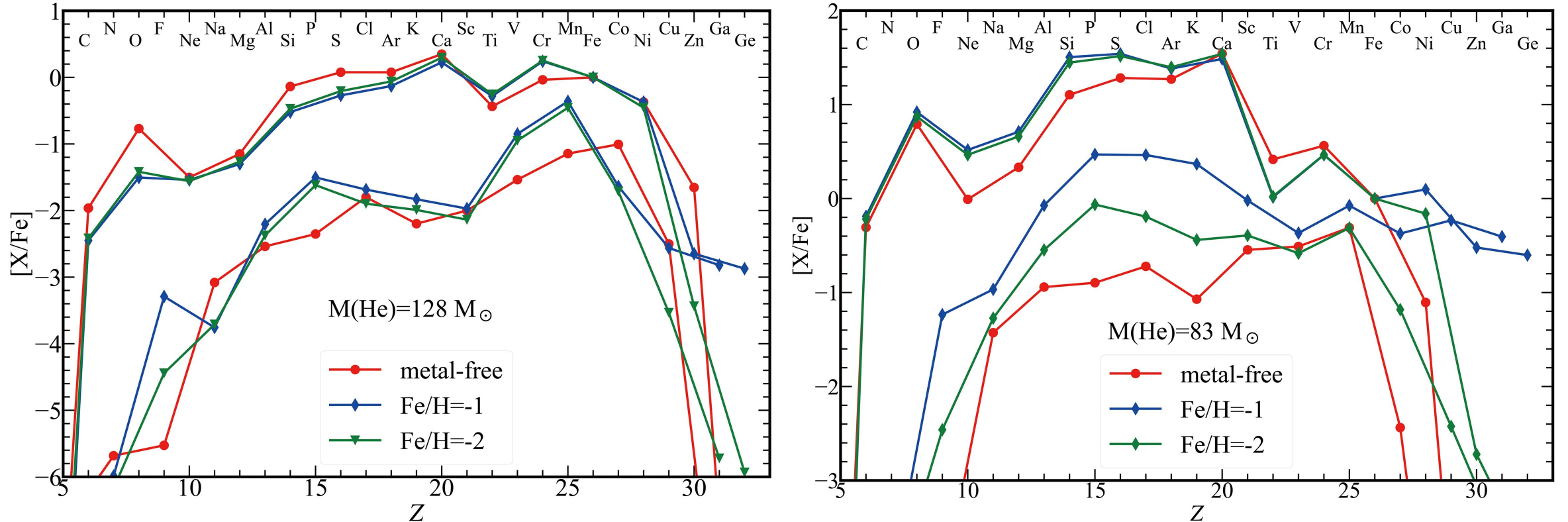}
\end{minipage}%
\caption{Abundance patterns of PISN yields normalized by iron yield
for stars with $M(\rm He)$ = 128 M$_\odot$ (left) and 83 M$_\odot$ (right).
\label{fig:pisn_metal}}
\end{figure*}

As shown in Figure \ref{fig:pisn_metal},
the odd-even effects in PISNe with [Fe/H] = $-$1 and $-$2 are weaker
than those in their metal-free counterparts.
For most elemental abundances,
predictions from the [Fe/H] = $-$2 models lie intermediate between those of
the [Fe/H] = $-$1 and metal-free models. However, they more closely approximate
the [Fe/H] = $-$1 results, owing to extensive convective mixing occurring
during the explosion phase in both cases.

In the He128 models, Pop II PISNe predict lower yields of even-$Z$ intermediate-mass elements
(particularly O, Si, and S) but higher yields of Fe-peak elements (such as Cr and Fe)
compared to the metal-free cases.
This pattern arises because convection transports light elements into hotter inner regions
while conveying Fe-peak nuclei outward to cooler layers (see Figure \ref{fig:isos_evo}).
The enhanced abundances of odd-$Z$ elements are a natural consequence of both the
higher initial metallicity and the resulting larger neutron excess.

In the He83 models, which explode before core Si burning, the abundances of 
even-$Z$ elements from O to Ar are enhanced by less than 0.5 dex in Pop II PISNe.
Intermediate-mass elements therefore dominate the stellar yields,
while Fe yields decrease, which is mainly derived from the decay of $^{56}$Ni.
All the odd-$Z$ elements show increased abundances,
with notable enhancement in [Al/Fe], [P/Fe], [Cl/Fe], and
[K/Fe] of approximately 0.8, 1.4, 1.3, and 1.5 dex, respectively.

In the models with [Fe/H] $\leq-$2, [Ga/Fe] and [Ge/Fe] are
extremely low and fall below detectable levels.
However, in the models with [Fe/H] = $-$1,
both [Ga/Fe] and [Ge/Fe] predicted by He128 models exceed -3 dex,
and those predicted by He83 exceed -1 dex.
The sensitivity of their yields to initial metallicity offers a
preliminary indication of weak s-process occurring.
The weak odd-even effect and the enhancements of Zn-Ge elemental yields are
the characteristic features to distinguish Pop II PISNe from Pop III PISNe.

\subsection{Comparisons and Discussions} \label{sec:diss}

\begin{figure}[H]
\centering
\begin{minipage}[c]{0.48\textwidth}
\includegraphics [width=85mm]{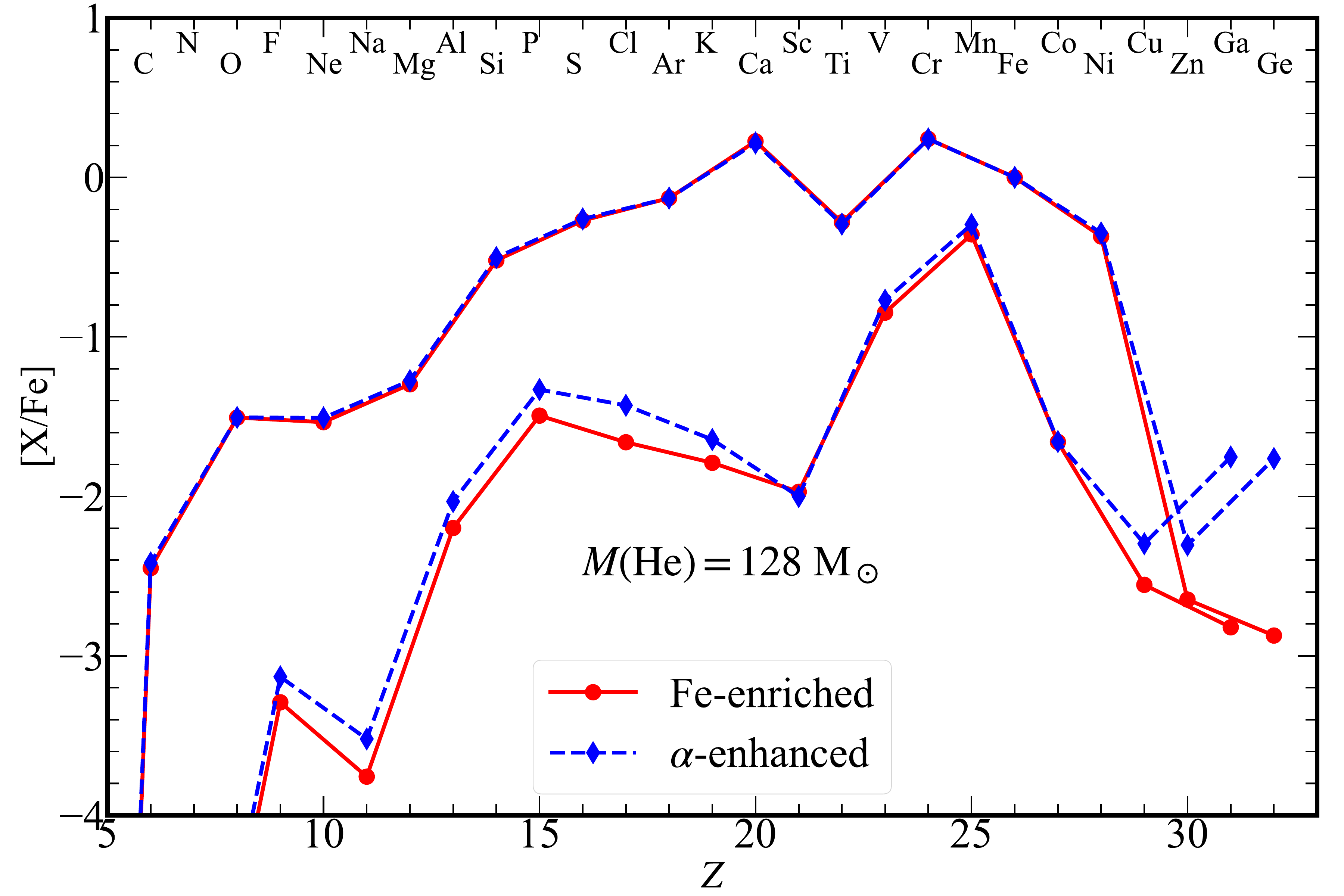}
\end{minipage}%
\caption{Abundance patterns of PISN yields normalized by iron
yield predicted by stars with [Fe/H] = -1 for different initial
compositions.
The red line represents the results using the Fe-enriched composition,
which is only enriched by Pop III PISNe.
While the blue line represents the composition revised by observations
from ref. \cite{2018ApJS..237...13L}, which is mainly enriched by CCSN.
\label{fig:pisn_ini}}
\end{figure}

We have previously emphasized that the Fe-enriched composition
represents an extreme limiting case. Therefore, it is necessary to explore
how different initial compositions impact the predicted elemental abundances.
In Figure \ref{fig:pisn_ini},
we present an additional test using the $M(\rm He)$ = 128 M$_\odot$ model
at [Fe/H] = -1, to quantify the impact of different initial compositions on
the final yields.
The ``Fe-enriched'' represents a composition only enriched by Pop III PISNe,
while the ``$\alpha$-enhanced''  provides a more realistic alternative,
enriched by multiple Pop III sources, particularly for CCSNe.

We find that the predicted abundances of even-Z elements
and Fe-peak elements are largely insensitive to the initial compositions.
This implies that variations in the initial composition have a negligible
impact on the stellar evolutionary tracks and explosion properties.
However, the Fe-enriched model predicts slightly lower yields of odd-$Z$
elements (e.g., Na, Al, P, Cl, and K), which is attributed to the lower
initial neutron excess in this scenario.
In total, the abundance deviations for elements from C to Ni across
different initial composition models are less than 0.2 dex.
The most pronounced differences occur for Cu, Zn, Ga, and Ge;
Notably, the Ga and Ge abundances in the ``Fe-enriched'' model are
lower than those in the ``$\alpha$-enhanced'' models by more than
one order of magnitude. This discrepancy arises 
because our Fe-enriched composition is neutron-deficient,
with a $^{14}$N mass fraction approximately five orders of magnitude
lower than that in the ``$\alpha$-enhanced'' composition as shown in
Figure \ref{fig:solar_ratio}.

\begin{figure}[H]
\centering
\begin{minipage}[c]{0.48\textwidth}
\includegraphics [width=85mm]{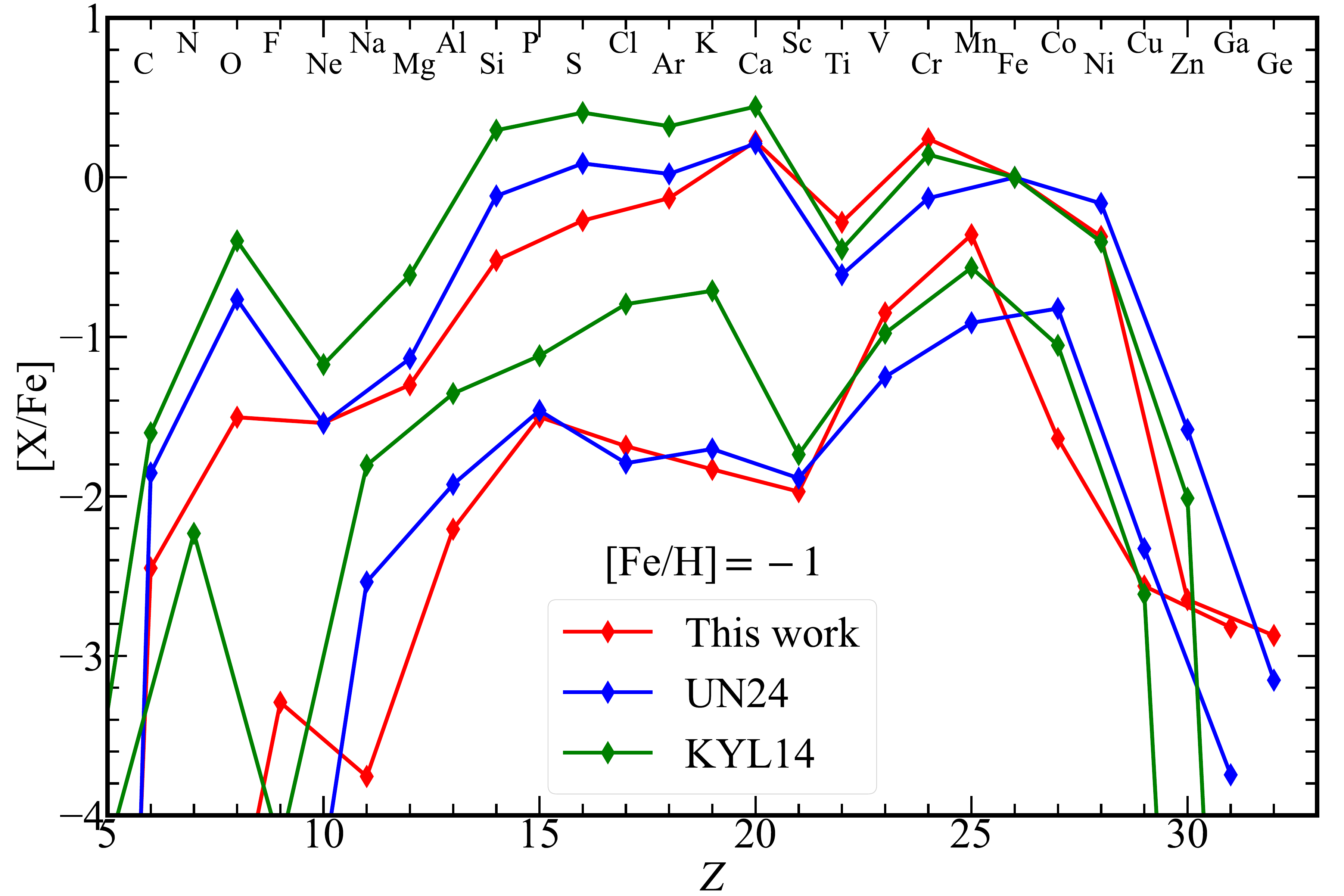}
\end{minipage}%
\caption{Abundance patterns of PISN yields normalized by iron
yield for stars with [Fe/H] = -1 for different studies.
The red lines are the results of this study.
The UN24 (blue) and KYL14 (green) represent the results from refs.
\cite{2024ApJ...961..146U} and \cite{2014AA...566A.146K}.
The He (CO) core masses for these models are 128 (116) M$_\odot$,
112 (112) M$_\odot$ and 121 (110) M$_\odot$, respectively. 
\label{fig:pisn_ref}}
\end{figure}

Figure \ref{fig:pisn_ref} compares the abundances predicted in this study
with those from KYL14 and UN24 for models with [Fe/H] = -1.
The ZAMS masses in these three studies are 244, 250, and 330 M$_\odot$,
corresponding to He (CO) core masses of 128 (116), 121 (110), and 112.5
(112.5) M$_\odot$, respectively. UN24 adopted higher initial masses
primarily due to a stronger mass-loss prescription,
Consequently, the entire H-envelope and He shells are stripped.
The He core mass in KYL14 is slightly smaller than ours,
mainly because core overshooting was not included.
The initial metallicities used in these three studies
are $Z$ = 1.86$\times 10^{-3}$, 0.001, and 0.1$Z_\odot$, respectively.
In this study, we adopt a Fe-enriched composition,
while the other two studies adopted solar composition.

Although the models in the other two studies adopted metallicities
close to [Fe/H] = $-$1, the predicted properties appear to be more
consistent with those of metal-free stars.
For instance, KYL14 reported a maximum central temperature of $10^{9.71}$ K,
a central density of 10$^{6.69}$ g cm$^{-3}$ and a $^{56}$Ni mass of 19.3 M$_\odot$,
respectively, which are very comparable to our metal-free model with $M(\rm He)=$ 
113 M$_\odot$ and the metal-free model in HW02 with $M(\rm He)=$ 115 M$_\odot$.
Similarly, UN24's model predicted a maximum central temperature of
$10^{9.80}$ K, a total energy of $E_{\rm tot}$ = 6.73 $\times 10^{52}$ erg,
and a $^{56}$Ni mass of $M(^{56}$Ni) = 36.6 M$_\odot$.
Our metal-free model with $M(\rm He)=$ 128 M$_\odot$ and
HW02's metal-free model with $M(\rm He)=$ 130 M$_\odot$
also show comparable maximum central temperatures and $^{56}$Ni yields
but slightly higher total energies, as summarized in Table \ref{tab:my-pisn}
in this work and in Table 2 from the ref. \cite{2002ApJ...567..532H}.

As shown in Figure \ref{fig:pisn_ini}, the main difference between
the initial compositions is the neutron excess, which modulates the yields
of odd-$Z$ elements and significantly affects Cu--Ge production.
While the initial neutron excess in KYL14 is similar to ours
($\eta_{\rm ini} \sim 10^{-4}$), their maximum post-explosion neutron excess
(1.6$\times10^{-3}$) is approximately five times lower than ours.
We attribute this to convection during the explosion, which 
significantly boosts the neutron excess (see Section \ref{sec:evo_exp_popII}),
thereby overwhelming the impact of the initial composition.
Additionally, since the CO core masses across the three models vary by
less than 5\%, the effect of core mass on nucleosynthesis is negligible
relative to the dominant role of explosive convection.
The absence of explosive convection in both studies leads to lower
neutron excesses and peak temperatures, resulting in reduced Fe-peak
element yields (particularly Fe)
and consequently elevated [X/Fe] ratios for elements lighter than Ca.

Although UN24 investigated the effects of rotation on Pop II PISNe,
their study primarily focused on the rotationally enhanced wind mass loss.
However, rotation also significantly enhances convection and internal mixing,
particularly during the explosive phase.
This may further enhance the neutron excess.
We expect that the yields of Zn--Ge and s-process elements should be further
enhanced by considering the rotation. A detailed investigation of these effects
is therefore a crucial direction for future research.

In this study, only the effect of $^{12}$C$(\alpha, \gamma)^{16}$O and
$^{16}$O+$^{16}$O reaction rates are investigated.
However, other reactions like 3$\alpha$, $^{12}$C+$^{12}$C, $^{22}$Ne+$\alpha$
and $^{17}$O+$\alpha$ also play crucial roles in determining the final
nucleosynthetic yields. Particularly, the update of $^{22}$Ne+$\alpha$
and $^{17}$O+$\alpha$ may affect the production of potential weak s-process
isotopes \cite{2026NuScT..37..146X}. 
While the 1D TDC model used in our calculation can handle
rapid convective growth on dynamic timescales better than standard MLT,
a 1D treatment may still artificially smooth or accelerate mixing over a short
blastwave timescales. In the forthcoming study, full 3D hydrodynamic calculations
will be necessary to quantitatively confirm the enhanced mixing efficiencies
presented here.

\subsection{The Implications for Chemical Enrichment} \label{sec:CE}

\begin{figure*}[]
\centering
\begin{minipage}[c]{0.9\textwidth}
\includegraphics [width=160mm]{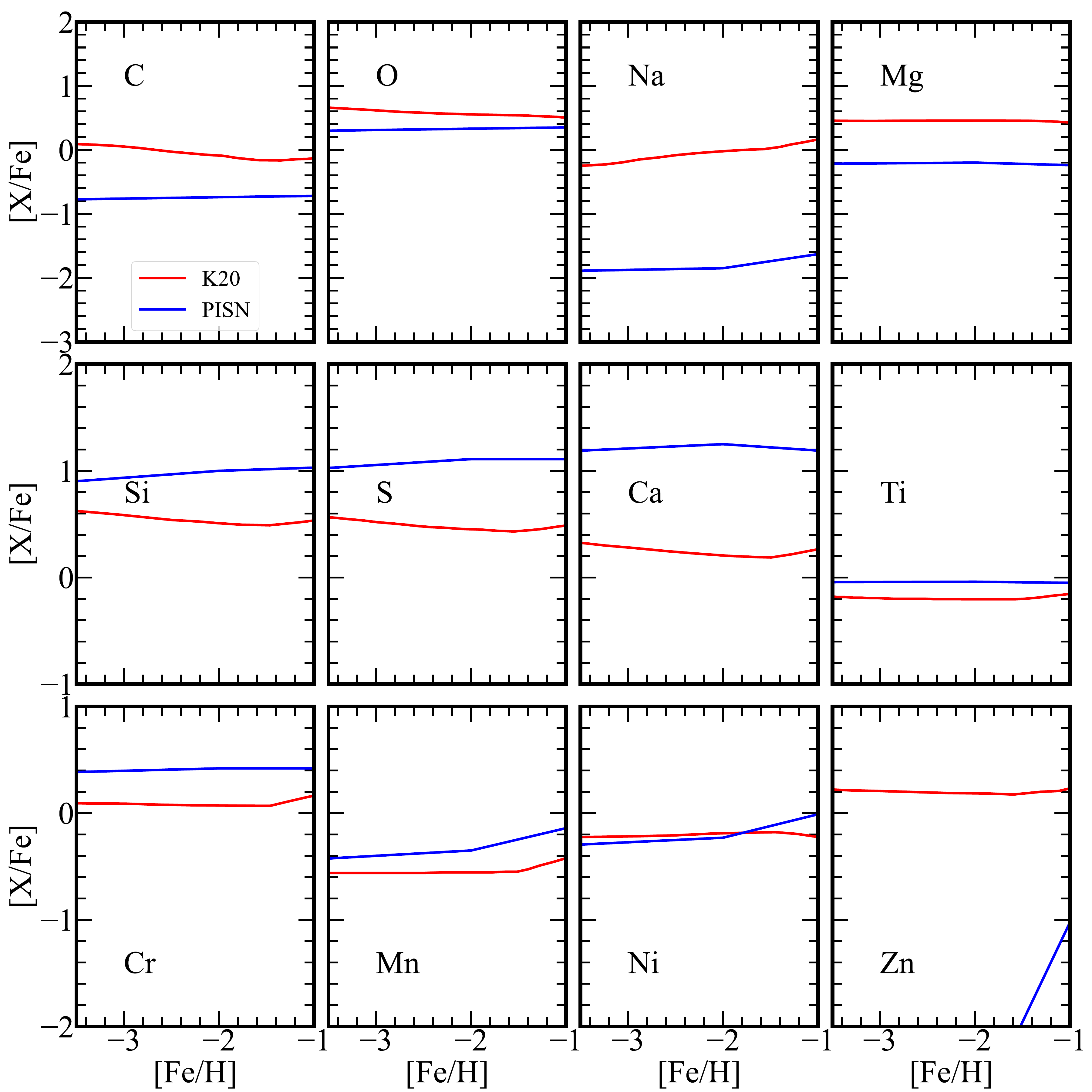}
\end{minipage}%
\caption{The evolution of the representative elemental abundances [X/Fe]
as a function of [Fe/H] predicted by PISNe (blue) in this work and
CCSNe+SN Ia (red) from \cite{Kobayashi_2020}.
\label{fig:pisn_enrich}}
\end{figure*}

As previously discussed, the enrichment of primordial gas by Pop~III PISNe is
expected to enhance the abundances of $\alpha$ isotopes relative to
their solar values, while substantially suppressing neutron-rich isotopes.
Consequently, the inclusion of PISN contributions is likely to
alter predictions from Galactic Chemical Evolution (GCE) models.
The magnitude of this deviation depends heavily on the birth rate of PISNe
in the early Universe. Given that the IMF of the early universe remains uncertain,
we adopt an extreme scenario in which chemical enrichment is driven solely by PISNe.

Using the Pop~II PISN models presented in Table~\ref{tab:my-pisn},
we interpolate and extrapolate the yields to cover the complete
PISN mass range from $M(\rm He)$ = 73 to 133 M$_\odot$.
The derived abundances are integrated over a Salpeter IMF
with a slope of $\gamma = -2.35$.
We should note that the PISNe yields vary nonlinearly with core mass,
particularly for Fe-peak elements and $^{56}$Ni production.
According to the discussion from ref. \cite{2014AA...566A.146K}
the linear interpolation typically introduces an uncertainty of $\sim$0.1--0.2 dex.
Therefore, the results presented in this section should be interpreted
as qualitative trends regarding the chemical signatures of PISNe,
rather than precise quantitative predictions.

Figure~\ref{fig:pisn_enrich} presents the changes of the representative elemental
abundances in a PISN-only enrichment scenario as a function of [Fe/H],
contrasted with the predictions from ref. \cite{Kobayashi_2020} that incorporate
conventional supernovae models (CCSN+SN Ia).
It is important to note that in the early Universe ([Fe/H] $< -1$),
CCSNe are the primary source of chemical enrichment for these elements.

The most notable characteristic of PISN-dominated enrichment is
the pronounced enhancement in the abundances of $\alpha$-elements from Si to Ca.
In contrast, abundances of lighter $\alpha$-elements (like C, O, and Mg) tend
to be lower than those predicted in ref. \cite{Kobayashi_2020} based on the
conventional supernovae models.
The odd-Z elements (Na) show significant deficits of more than 1.5 dex in 
PISN-only enrichment scenarios,
which implies that the CCSNe are the primary sources of odd-Z elements.
Zn shows a significant sensitivity to metallicity
because they produced mainly by the s-process in PISN-only enrichment scenarios.
Pop II PISNe may represent one of the potential sources of heavy elements
in the early universe. As discussed in ref. \cite{2005IAUS..228..297H},
the production of s-process isotopes may be enhanced
if some $^{14}$N is mixed into the He core before the onset of
pair-instability.
Such mixing processes might result from rotation or varying degrees of
convective overshooting.
In our forthcoming studies, we will discuss the s-process in Pop II PISNe
in detail.

\section{Summaries} \label{sec:summary}

In this paper, we have conducted a systematic study of PISN
using the \texttt{MESA} code. We calculated a grid of models ranging
from primordial Pop III stars to Fe-enriched Pop II stars,
focusing on the effects of the key nuclear reaction rate uncertainties
and explosive convective mixing.
Our findings provide crucial insights for identifying PISN imprints
in metal-poor stars and understanding early chemical enrichment.
We summarize our results as follows:

\textbf{1. Evolution and Nucleosynthesis of Pop III PISNe}

We calculated a series of Pop III PISN models with initial masses of
$M(\rm ZAMS)$ = 130 - 300 M$_\odot$.
By varying the $^{12}$C$(\alpha, \gamma)^{16}$O and $^{16}$O+$^{16}$O
reaction rates, we found that:

\noindent (1) Increasing the $^{12}$C$(\alpha, \gamma)^{16}$O rate by 1 $\sigma$
or decreasing the $^{16}$O+$^{16}$O rate by a factor of 10
leads to explosions at higher central temperatures.
This results in higher total energies and increased production of $^{56}$Ni.
The underlying mechanisms differ: a higher $^{12}$C$(\alpha, \gamma)^{16}$O
rate reduces the core carbon fraction,
weakening shell C-burning and leading to denser cores.
Conversely, a lower $^{16}$O+$^{16}$O delays O-burning ignition,
requiring further contraction.

\noindent (2) Compared to previous studies (e.g., HW02, TYU18),
our models show that odd-$Z$ elements (e.g., Na, Al, Cl, K) and Fe-peak elements
are sensitive to the $^{12}$C$(\alpha, \gamma)^{16}$O rate,
while only Fe-peak elements are sensitive to the $^{16}$O+$^{16}$O rate.
The predicted abundance ratios of odd-$Z$ elements in this work generally
fall between those of previous studies (HW02, TYU18).

\textbf{2. Evolution and Nucleosynthesis of Fe-enriched Pop II PISNe}

We investigated Pop II PISNe born in environments polluted solely
by Pop III PISNe, assuming a galaxy where massive Pop II stars form
promptly in a highly localized region enriched solely by Pop III PISNe,
before other supernova products can mix in.
The Fe-enriched Pop II PISNe are simulated with $M(\rm He)$ = 83 and 128
M$_\odot$ and initial metallicities of [Fe/H] = -2 and -1.

\noindent (1) Unlike previous works that neglected convection during explosion,
we find that the higher opacity in metal-enriched models triggers
vigorous convective mixing during explosive O-burning.
This explosive convection significantly enhances energy transport and core contraction.
Consequently, the stars explode at higher temperatures and densities.
This process transports more $^{4}$He into the core to be consumed,
markedly increasing the explosion energy and the yields of $^{28}$Si
(in lower-mass models) or $^{56}$Ni (in higher-mass models).

\noindent (2) We explored the chemical evolution under a PISN-only enrichment scenario.
Compared to conventional GCE models (CCSNe + SNe Ia),
a PISN-dominated environment shows distinct features:
(a) significant enhancement of $\alpha$-elements from Si to Ca, and
(b) severe depletion of odd-$Z$ elements by more than 1.5 dex.
We also identified the production of potential $s$-process elements (Zn, Ga, Ge)
in [Fe/H] = $-1$ models, suggesting Pop II PISNe could contribute to the early
chemical inventory of heavy elements.
Given that the nuclear network in this work is
truncated at Ge ($Z=32$), future studies should incorporate
an extended network covering the light s-process peak.

\Acknowledgements{The authors would like to thank the referees for their
constructive comments and suggestions.
This work is supported by the National Natural Science
Foundation of China under Grant Nos. 12588202, 12473028, 12541303, 12090040,
and 12090042.
W. Y. X. is supported by the Cultivation Project for LAMOST Scientific Payoff
and Research Achievement.
K. N. is supported by the World Premier International Research Center Initiative
(WPI), MEXT, Japan, and the Japan Society for the Promotion of Science
JSPS KAKENHI Grant Numbers JP21H044pp, JP23K03452, and JP25K01046.
Q-F, X. is supported by grants Nos.12422304 from the National Natural Science
Foundation of China grant and grant No. XDB1160301 from the Strategic Priority
Research Program of Chinese Academy of Sciences.
C. M. Y. is supported by grants from the Research Grant Council of the Hong Kong
Special Administrative Region, China (Project Nos. 14300320 and 14304322) and
the European Union through ERC Synergy Grant HeavyMetal no. 101071865.}

\InterestConflict{The authors declare that they have no conflict of interest.}



\bibliographystyle{scpma}

\begin{appendix}

\section{Comparison with Previous Studies}

As we have mentioned, central $X$($^{12}$C) is an important quantity,
which affects the final explosion of massive stars.
Thus, additional comparison of $X$($^{12}$C) in different studies
is discussed here.

In Figure \ref{fig:xc_sigma}, we have shown central $X$($^{12}$C)
at the end of core He burning. 
HW02 adopted the $^{12}$C$(\alpha, \gamma)^{16}$O rate from ref.
\cite{1996ApJ...468L.127B}, which is slightly higher than our fiducial value.
Yet, they reports a higher $X$($^{12}$C) at He exhaustion.
TYU18 adopted the CF88 rate with $f_{\rm CF88}$ from 0.1 to 1.2,
which is lower than those used in this study.
A representative model with $M(\rm ZAMS)$ = 180 M$_\odot$ and
$f_{\rm CF88}$ = 1.2 reported in TYU18 predicts $X$($^{12}$C) of 0.144.
This implies that all the models calculated in TYU18 predicted a higher
$X$($^{12}$C) than those in this study.
These discrepancies could arise from their use of a different $3\alpha$ rate
or from differences in the treatment of convection mixing.

\begin{figure}[H]
\centering
\begin{minipage}[c]{0.48\textwidth}
\includegraphics [width=82mm]{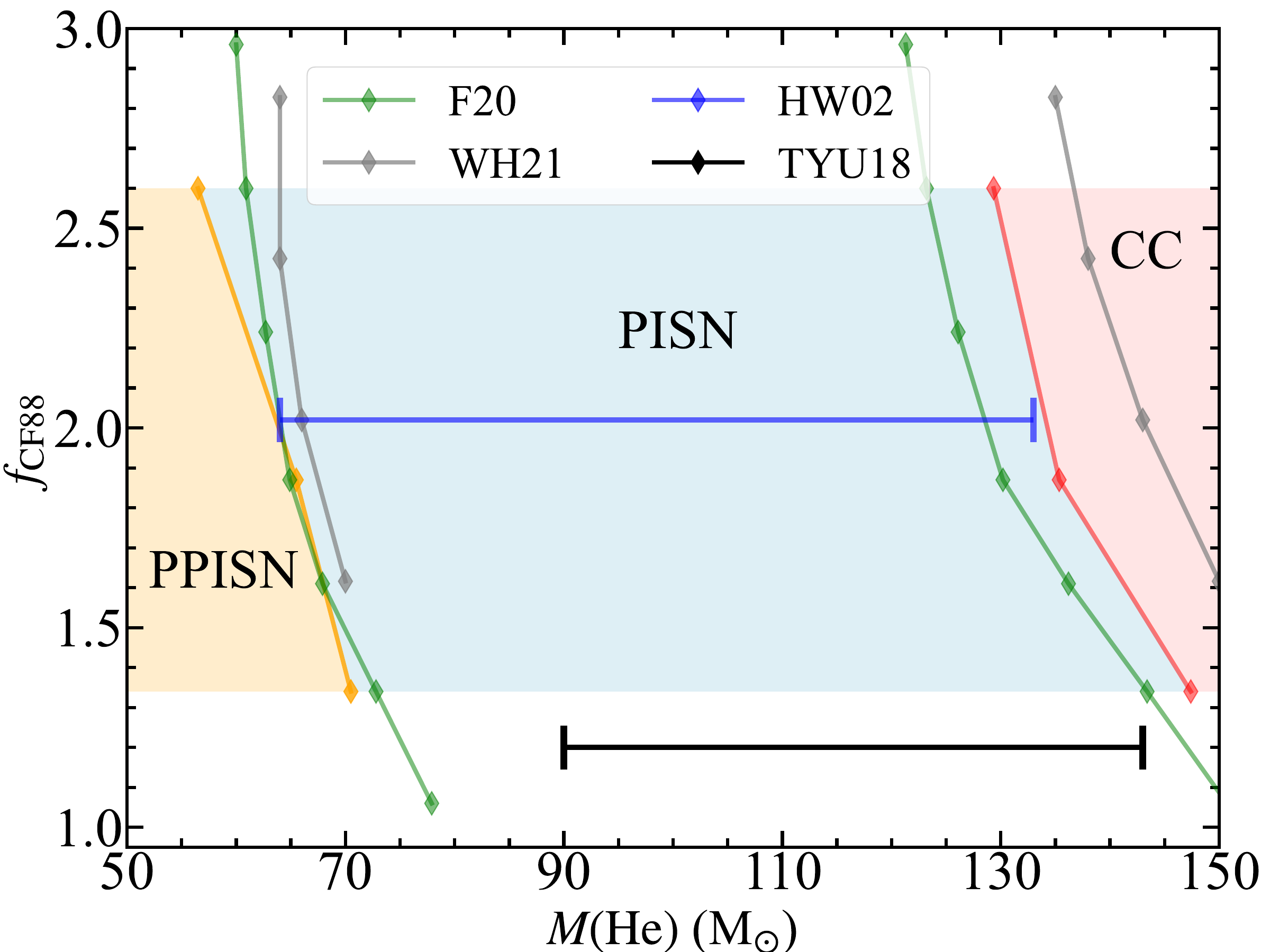}
\end{minipage}%
\caption{The comparison of PPISN/PISN and PISN/CC boundaries between this work
and other works. All the rates of $^{12}$C$(\alpha,\gamma)^{16}$O used in different
works are normalized by approximately multipliers on the CF88 rate.
The green and gray lines show the boundaries predicted by F20
\cite{Farmer_2020} and WH21 \cite{2021ApJ...912L..31W}, 
respectively. The blue and black lines show the boundaries from HW02 and TYU18,
respectively.
\label{fig:pisn}}
\end{figure}

As a result, different $X$($^{12}$C) shifts the explosion results of
the massive progenitors, in Figure \ref{fig:pisn},
we compare the boundaries between PPISN/PISN and PISN/CC
predicted in this study with those from previous studies
\cite{Farmer_2020, 2021ApJ...912L..31W, 2002ApJ...567..532H,
Takahashi_2018_1, 2026arXiv260319883X}.

The stars in this work explode as PISNe ranging
from $M(\rm He)$ = 57.8 - 128.2 M$_\odot$, 62.6 - 133.5 M$_\odot$
and 72.5 - 143.2 M$_\odot$ for $\sigma_{C12\alpha}$ = 1, 0 and $-$1
($f_{\rm CF88} \simeq$ 2.60, 1.87 and 1.34), respectively.
This result agrees well with HW02.
F20 and WH21 predicted different boundaries mainly because they adopted $Z=10^{-3}$
instead of metal-free. More details about the effect of physical parameters on
PPISN/PISN boundaries have been discussed by previous works
\cite {Farmer_2019, 2021ApJ...912L..31W}.
These boundaries reported by TYU18 shift to higher $M(\rm He)$ due to low $f_{\rm CF88}$.
Because they include an H-rich envelope in the models, the gravitational energies 
for the same He core masses should be larger than those in HW02 and this work.
As a result, the mass range predicted in TYU18 is narrower.

\section{Abundance Patterns} \label{sec:abund}

\begin{figure}[H]
\centering
\begin{minipage}[c]{0.48\textwidth}
\includegraphics [width=85mm]{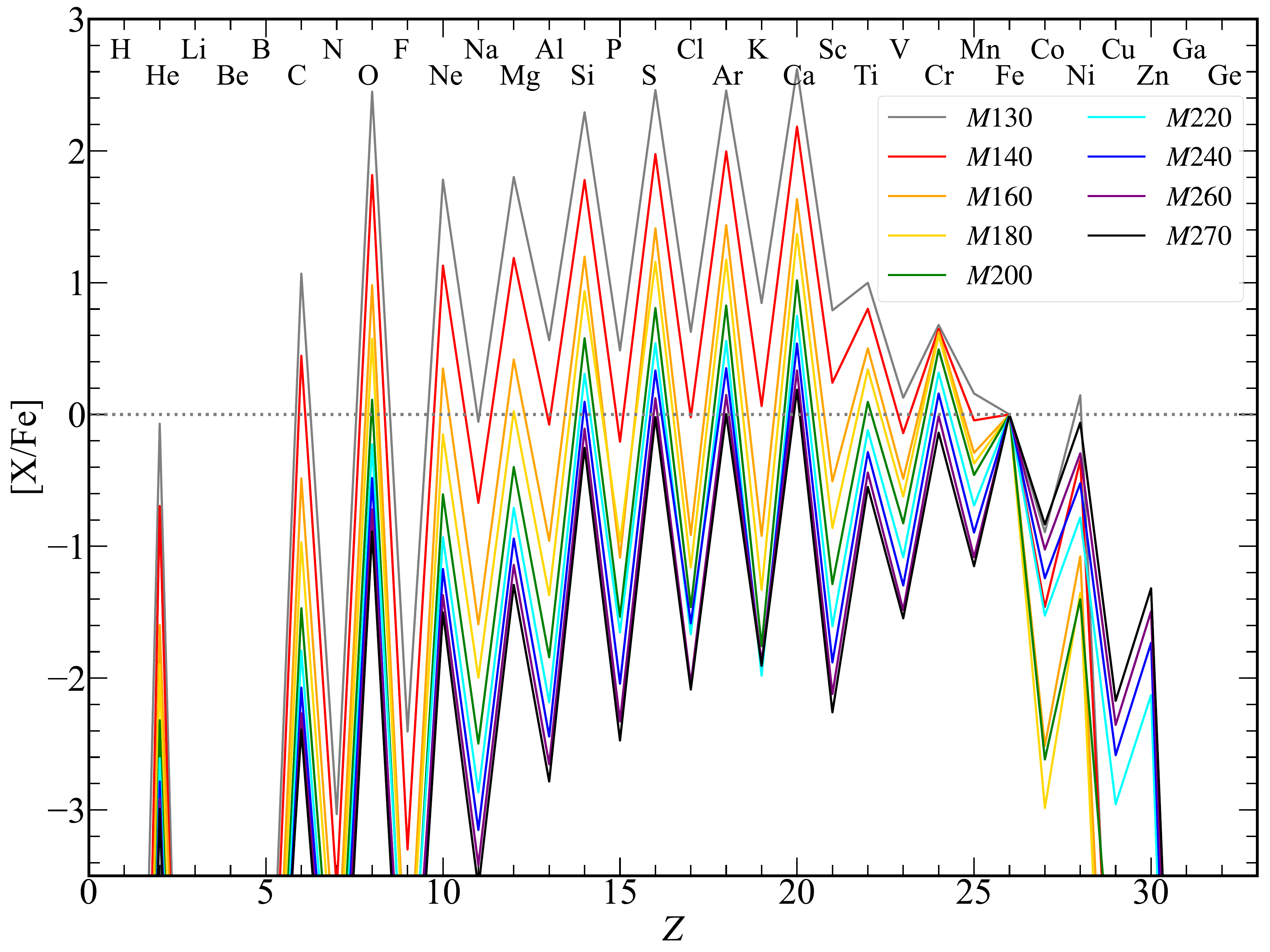}
\end{minipage}%

\begin{minipage}[c]{0.48\textwidth}
\includegraphics [width=85mm]{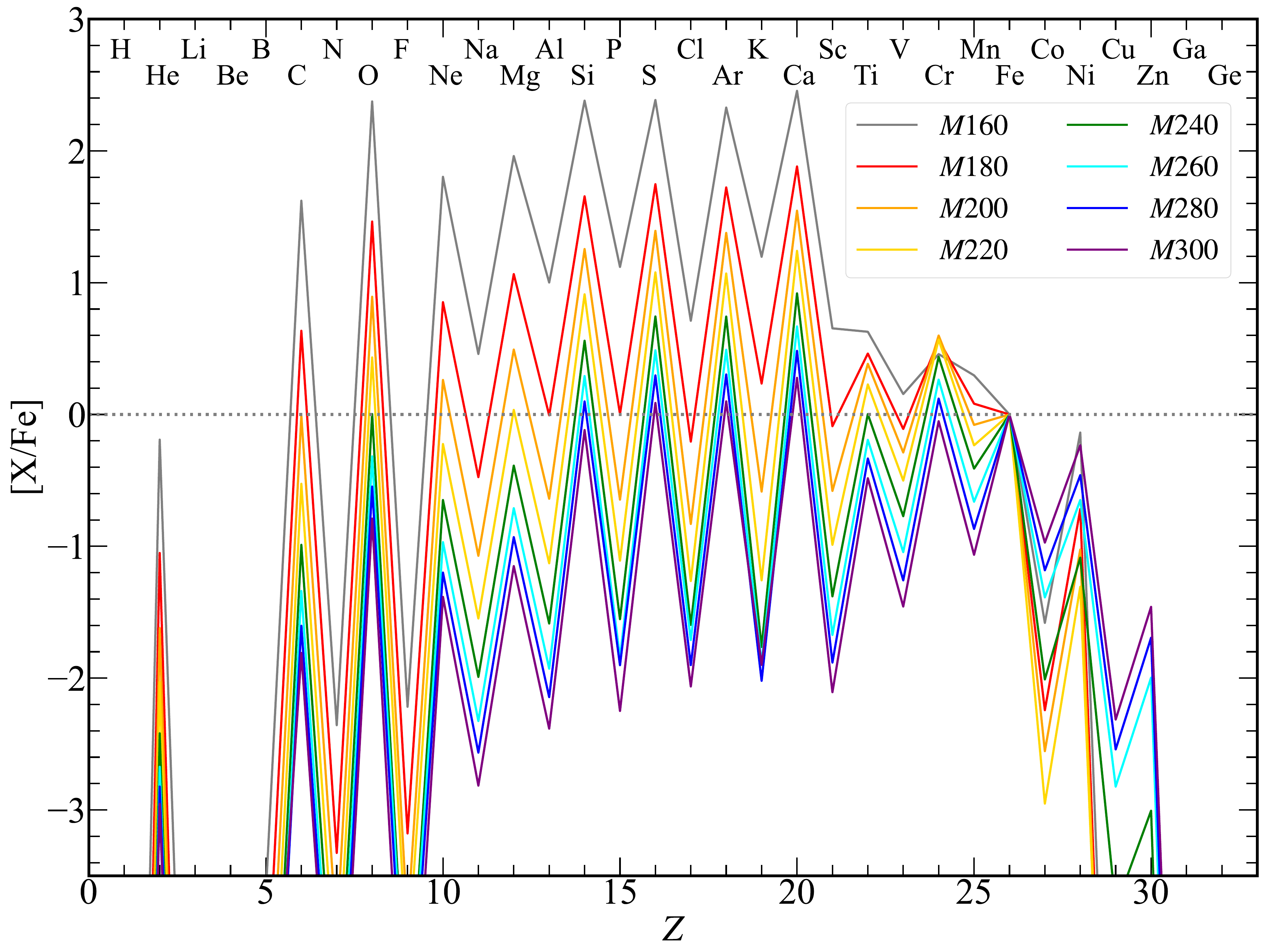}
\end{minipage}%
\caption{Abundance patterns of PISN yields for $\sigma_{C12\alpha}$ = 1 (upper) and -1 (bottom).
\label{fig:pisn_yield000}}
\end{figure}

In this section, Figure \ref{fig:pisn_yield000} presents the abundance patterns of
Pop III PISNe with $\sigma_{C12\alpha}$ = $\pm1$, as a supplemental comparison of
Figures \ref{fig:pisn_yield} and \ref{fig:pisn_sigma}.

\section{Initial Composition of Fe-enriched Pop II PISNe} \label{tab:yield2}

The initial composition is calculated by assuming [Fe/H] = 0
in the local ISM after Pop III PISNe explosion according to Equation \ref{equ:fe_h}.
Because almost no s- and r-process isotopes are produced in the Pop III PISNe,
the mass fraction of the elements heavier than Ge is not included in this composition.

\begin{table*}[t]
\centering
\caption{This composition is scaled to [Fe/H] = 0 with $Z= 0.0186$.}
\label{tab:my-solar}
\begin{tabular}{cccccc}
\toprule
Isotopes  & Mass fraction  &  Isotopes  & Mass fraction &  Isotopes  & Mass fraction  \\
\midrule
$^{1}$H   & 7.3681 $\times 10^{-01}$ & $^{31}$P  & 2.4500 $\times 10^{-07}$ & $^{52}$Cr & 3.2485 $\times 10^{-05}$ \\ 
$^{2}$H   & 3.9345 $\times 10^{-05}$ & $^{32}$S  & 1.7832 $\times 10^{-03}$ & $^{53}$Cr & 1.0605 $\times 10^{-06}$ \\  
$^{3}$He  & 2.3430 $\times 10^{-05}$ & $^{33}$S  & 1.2700 $\times 10^{-06}$ & $^{54}$Cr & 1.4735 $\times 10^{-13}$ \\   
$^{4}$He  & 2.4489 $\times 10^{-01}$ & $^{34}$S  & 1.3078 $\times 10^{-07}$ & $^{55}$Mn & 2.5688 $\times 10^{-06}$ \\ 
$^{6}$Li  & 4.5467 $\times 10^{-21}$ & $^{36}$S  & 6.1969 $\times 10^{-13}$ & $^{54}$Fe & 8.6001 $\times 10^{-06}$ \\ 
$^{7}$Li  & 2.2174 $\times 10^{-09}$ & $^{35}$Cl & 1.3134 $\times 10^{-07}$ & $^{56}$Fe & 1.3269 $\times 10^{-03}$ \\ 
$^{9}$Be  & 3.1223 $\times 10^{-23}$ & $^{37}$Cl & 2.0771 $\times 10^{-07}$ & $^{57}$Fe & 1.3266 $\times 10^{-05}$ \\   
$^{10}$B  & 5.2650 $\times 10^{-20}$ & $^{36}$Ar & 4.3345 $\times 10^{-04}$ & $^{58}$Fe & 9.7058 $\times 10^{-13}$ \\  
$^{11}$B  & 1.5140 $\times 10^{-15}$ & $^{38}$Ar & 1.2883 $\times 10^{-07}$ & $^{59}$Co & 2.2874 $\times 10^{-07}$ \\ 
$^{12}$C  & 2.9586 $\times 10^{-04}$ & $^{40}$Ar & 1.3460 $\times 10^{-14}$ & $^{58}$Ni & 1.2685 $\times 10^{-05}$ \\ 
$^{13}$C  & 9.5018 $\times 10^{-12}$ & $^{39}$K  & 8.0574 $\times 10^{-08}$ & $^{60}$Ni & 7.5678 $\times 10^{-06}$ \\ 
$^{14}$N  & 6.5489 $\times 10^{-09}$ & $^{40}$K  & 4.9338 $\times 10^{-12}$ & $^{61}$Ni & 2.5829 $\times 10^{-07}$ \\ 
$^{15}$N  & 4.9685 $\times 10^{-09}$ & $^{41}$K  & 7.0583 $\times 10^{-08}$ & $^{62}$Ni & 1.4320 $\times 10^{-06}$ \\  
$^{16}$O  & 1.0483 $\times 10^{-02}$ & $^{40}$Ca & 5.7915 $\times 10^{-04}$ & $^{64}$Ni & 6.8885 $\times 10^{-19}$ \\   
$^{17}$O  & 1.0957 $\times 10^{-10}$ & $^{42}$Ca & 4.3607 $\times 10^{-09}$ & $^{63}$Cu & 6.8675 $\times 10^{-10}$ \\  
$^{18}$O  & 1.9567 $\times 10^{-11}$ & $^{43}$Ca & 2.4833 $\times 10^{-10}$ & $^{65}$Cu & 1.0177 $\times 10^{-09}$ \\  
$^{19}$F  & 8.3111 $\times 10^{-12}$ & $^{44}$Ca & 2.5514 $\times 10^{-07}$ & $^{64}$Zn & 1.2840 $\times 10^{-08}$ \\  
$^{20}$Ne & 4.7052 $\times 10^{-04}$ & $^{46}$Ca & 1.9259 $\times 10^{-19}$ & $^{66}$Zn & 1.6349 $\times 10^{-08}$ \\  
$^{21}$Ne & 6.3121 $\times 10^{-09}$ & $^{48}$Ca & 1.9626 $\times 10^{-26}$ & $^{67}$Zn & 1.4306 $\times 10^{-11}$ \\   
$^{22}$Ne & 7.9672 $\times 10^{-09}$ & $^{45}$Sc & 2.8838 $\times 10^{-09}$ & $^{68}$Zn & 2.8363 $\times 10^{-12}$ \\  
$^{23}$Na & 3.2680 $\times 10^{-07}$ & $^{46}$Ti & 3.5233 $\times 10^{-09}$ & $^{70}$Zn & 2.8021 $\times 10^{-31}$ \\ 
$^{24}$Mg & 3.6345 $\times 10^{-04}$ & $^{47}$Ti & 2.8407 $\times 10^{-10}$ & $^{69}$Ga & 7.1613 $\times 10^{-17}$ \\ 
$^{25}$Mg & 7.1303 $\times 10^{-08}$ & $^{48}$Ti & 2.6996 $\times 10^{-06}$ & $^{71}$Ga & 2.2751 $\times 10^{-24}$ \\ 
$^{26}$Mg & 8.6955 $\times 10^{-08}$ & $^{49}$Ti & 7.5310 $\times 10^{-08}$ & $^{70}$Ge & 3.9393 $\times 10^{-19}$ \\ 
$^{27}$Al & 1.7986 $\times 10^{-06}$ & $^{50}$Ti & 4.5497 $\times 10^{-15}$ & $^{72}$Ge & 2.4173 $\times 10^{-26}$ \\  
$^{28}$Si & 2.4181 $\times 10^{-03}$ & $^{50}$V  & 0.0000 $\times 10^{+00}$ & $^{73}$Ge & 4.5060 $\times 10^{-31}$ \\
$^{29}$Si & 2.4761 $\times 10^{-06}$ & $^{51}$V  & 3.6408 $\times 10^{-08}$ & $^{74}$Ge & 0.0000 $\times 10^{+0}$  \\ 
$^{30}$Si & 2.2279 $\times 10^{-07}$ & $^{50}$Cr & 5.5499 $\times 10^{-08}$ &                             \\              
\bottomrule
\end{tabular}
\end{table*}

\renewcommand{\thesection}{Appendix}

\end{appendix}

\end{multicols}
\end{document}